\documentclass[11pt,a4paper]{article}
\author{}

\usepackage[utf8]{inputenc} 
\usepackage{amssymb, amsbsy} 
\usepackage{amsfonts}
\usepackage[centertags]{amsmath}
\usepackage{mathtools}
\usepackage{amsthm}
\usepackage{mathabx} 
\usepackage{nccmath}
\usepackage{array}
\usepackage{graphicx} 
\usepackage{subfigure}  
\usepackage[left=2cm,right=2cm,top=2cm,bottom=2cm]{geometry}
\usepackage{upgreek} 
\usepackage{cancel} 
\usepackage{mathdots} 
\usepackage{mathrsfs} 
\usepackage{stackrel} 
\usepackage{bm} 
\usepackage{newlfont}
\usepackage{fancyhdr}
\usepackage{float} 
\usepackage{subfigure}
\usepackage{longtable}
\usepackage{booktabs,makecell} 
\usepackage{slashbox}
\usepackage[dvipsnames]{xcolor}
\usepackage[final,pdftex,bookmarks=true,bookmarksopen=true,breaklinks=true,colorlinks=true,bookmarksnumbered=true,pdfsubject={ },pdfauthor={Ar\'is Fanjul Hevia}]{hyperref} 

\hypersetup{urlcolor=blue,linkcolor=blue,citecolor=blue}
\usepackage{natbib}
\usepackage{setspace}
\usepackage[toc,page]{appendix}
\usepackage{longtable}
\usepackage{multirow}
\usepackage{caption}
\usepackage{enumerate}
\usepackage{soul} 

\usepackage{authblk} 

\title{
{\textbf{A test for comparing conditional ROC curves with multidimensional covariates}}}

\author[1]{Ar\'is Fanjul-Hevia}
\author[2]{Juan Carlos Pardo-Fern\'andez}
\author[3]{Ingrid Van Keilegom}
\author[4]{Wenceslao Gonz\'alez-Manteiga}
\affil[1]{Departamento de Estad\'istica e Investigación Operativa y Didáctica de la Matemática, Universidad de Oviedo}
\affil[2]{Departamento de Estat\'istica e Investigaci\'on Operativa  and Centro de Investigaci\'ons Biom\'edicas (CINBIO), Universidade de Vigo}
\affil[3]{Research Centre for Operations Research and Statistics, KU Leuven}
\affil[4]{Departamento de Estat\'istica, An\'alise Matem\'atica e Optimizaci\'on, Universidade de Santiago de Compostela}
\date{}
\begin{document}

\maketitle

\definecolor{dgreen}{rgb}{0.,0.5,0.}

\renewcommand{\labelitemi}{$\sqbullet$}

%
\theoremstyle{plain}
\newtheorem{theorem}{Theorem}
\newtheorem{lemma}[theorem]{Lemma}
\newtheorem{prop}[theorem]{Proposition}
\newtheorem{cor}[theorem]{Corollary}
\theoremstyle{definition}
\newtheorem{mydef}{Definition}
\newtheorem{remark}{Remark}

\begin{abstract}
The comparison of Receiver Operating Characteristic (ROC) curves is  frequently used in the literature to compare the discriminatory capability of different classification procedures based on diagnostic variables. The performance of these variables can be  sometimes influenced by the presence of other covariates, and thus they should be taken into account when making the comparison. A new non-parametric test is proposed here for testing the equality of two or more dependent ROC curves conditioned to the value of a multidimensional covariate. Projections are used for transforming the problem into a one-dimensional approach easier to handle. Simulations are carried out to   study the practical performance of the new methodology. A real data set of patients with Pleural Effusion is analysed to illustrate this procedure.
\end{abstract}


\section{Introduction}
\label{sec1}

In any classification problem such as a diagnostic method --in which the aim is to discriminate between two populations, usually identified as the healthy population and the diseased population-- the main concern is to minimize the number of subjects that are misclassified.  Receiver Operating Characteristic (ROC) curves are commonly used in this context for studying the behaviour of the classification variables  \citep[see, for example, the monograph of][as an introduction to the topic]{Pepe2003}.  They combine the notions of sensitivity (the ability of classifying a diseased patient as diseased) and specificity (the ability of classifying a healthy individual as healthy), two measurements that can be expressed in terms of the cumulative distribution functions of the diagnostic variables of the diseased and the healthy populations. 

When there is more than one variable for diagnosing a certain disease one can compare their respective ROC curves in order to decide whether their discriminatory capability is different or not. 
 This is what happens in the medical example that we will be using in this paper for illustrating purposes, a real data set containing the information of patients with pleural effusion. In this data set there are two  variables (the carbohydrate antigen 152 and the cytokeratin fragment 21-1) that can be used for deciding whether that pleural effusion is due to the presence of a malignant tumour or not. The objective of the analysis will be to compare the diagnostic capability of those markers.

There are several methodologies discussed in the literature for making that sort of comparisons \citep[for a review of such methodologies, see][]{Fanjul-Hevia2018}, although most of them do not consider the possible effect that the presence of covariates can have in the performance of the test.  In the example provided, apart from the diagnostic variables there are other covariates such as the age or the neuron-specific enolase of the patients. It is important to take this information into account, because the diagnostic capability of a marker may change with the value of a covariate \citep{Pardo-Fernandez2014}. In this paper the aim is to propose a test to compare ROC curves that includes the presence of a multidimensional covariate in the analysis.

One way of introducing the effect of the covariates into the study is by using the \textit{conditional ROC curve}. If we consider $Y^F$ and $Y^G$ as the continuous diagnostic markers in the diseased and healthy populations, respectively,  $\bm{X}^F = {(X_1^F,\dots, X_d^F)}' $  as the continuous $d-$dimensional covariate of the diseased population and $\bm{X}^G = {(X_1^G,\dots, X_d^G)}' $ as the continuous $d-$dimensional covariate of the healthy population, then, given a fixed value $\bm{x} = {(x_1,\dots, x_d)}' \in \bm{R_{\bm{X}}}$ (where $\bm{R_{\bm{X}}}$ is the intersection of $\bm{R_{{X^F}}}$ and $\bm{R_{{X^G}}}$, 
the supports of $\bm{X}^F$ and $\bm{X}^G$, and is assumed to be non-empty), the conditional ROC curve is defined as 
\begin{eqnarray}
ROC^{\bm{x}}(p) = 1- F(G^{-1} (1-p|\bm{x})|\bm{x}), \;p \in (0,1),
\label{eq:defROCx}
\end{eqnarray}
where $F(y|\bm{x}) = P(Y^F\leq y | \bm{X}^F=\bm{x})$,  and $G(y|\bm{x}) = P(Y^G\leq y | \bm{X}^G=\bm{x})$.

By comparing these conditional ROC curves instead of the standard ROC curves it is possible to incorporate the potential effect of the covariates in the analysis of the equivalence of two or more methods of diagnosis. A test for performing this comparison is proposed in  \cite{Fanjul-Hevia2019} for the case of a continuous one-dimensional covariate. The objective here is to extend that methodology to the case in which we have a multidimensional covariate. Thus, the aim is to test, given  a certain $\bm{x} \in \bm{R_X}$,
\begin{eqnarray}
H_0: ROC_1^{\bm{x}}(p) = \dots = ROC_K^{\bm{x}}(p) \; \text{ for all } p \in (0,1),
\label{eq:H0X}
\end{eqnarray}
where $K$ is the number of diagnostic markers (and thus, ROC curves) that are being compared. In this context we would have $K$ diagnostic variables and one $d-$dimensional covariate in the healthy population, $(\bm{X}^F,Y_1^F,\cdots,Y_K^F)$, and similar variables in the diseased population, $(\bm{X}^G,Y_1^G,\cdots,Y_K^G)$. 
 In practice this kind of test could help to design a more personalised diagnostic method based on the covariate values of each patient.
With this methodology, in the medical example at hand we could determine whether the carbohydrate antigen 152 and the cytokeratin fragment 21-1 are equally suitable for the diagnosis of a patient with a certain age and a certain enolase value.

 In order to be able to make this comparison, we are going to rely on the estimation of the corresponding conditional ROC curves. There is a wide range of estimation methods in the literature: some of them estimate the conditional distribution functions involved in the definition of the conditional ROC curve, others use regression functions to include the effect of the covariates (following  direct or indirect approaches). 
 See \cite{Pardo-Fernandez2014} for a further review of this topic. 
 
 In  \cite{Fanjul-Hevia2019} the estimation of the conditional ROC curve that is used is based on the indirect (or induced) regression methodology, which incorporates the covariate information through regression models by considering the effect of those covariates in the diagnostic marker in each population of healthy or diseased separately. However, this method was originally designed for one single covariate. One could think of extending that methodology by changing the estimator of the conditional ROC curve for another capable of handling multidimensional covariates. Nevertheless, there are not many methods in the literature capable of considering more than one covariate when estimating the conditional ROC curve, and most of them have some parametric assumptions that we would like to avoid making. See \cite{Carvalho2013} as an example of a non-parametric Bayesian model to estimate the conditional distribution functions involved in the ROC curves, \cite{Rodriguez-Alvarez2011a} or \cite{Rodriguez-Alvarez2018} as an example of a direct ROC regression model or \cite{RodriguezMartinez2014} as an example of induced methodology (framed in a Bayesian setting). In our case we will be following a frequentist approach.



The tests related to multidimensional data tend to become less powerful when the dimension of the problem increases. This is why, in this paper, the problem of comparing conditional ROC curves is first transformed using projections in such a way that the multidimensional problem becomes a unidimensional problem easier to handle. This idea has been applied several times in the literature  for reducing the dimension in goodness-of-fit problems \citep[see, for example,][]{Escanciano2006,GarciaPortugues2014,Patilea2016}, but, to the best of our knowledge, it is the first time that it is applied on an ROC curve setting. In the last few years random projections are increasingly being used as a way to overcome the curse of dimensionality.
 The characterization  of the multidimensional distribution of the original data by the distribution of the randomly projected unidimensional data is what allows for the reduction of the dimension.

 To that end, in Section \ref{sec2} we show how (\ref{eq:H0X}) can be transformed in a test with one-dimensional covariates by using projections. Then, a methodology is proposed for testing that equivalent hypothesis.  In Section \ref{sec3} the results from a simulation study show the practical performance of the test in terms of level approximation and power. 
 The procedure is illustrated in Section \ref{sec4} by analysing the real data set containing information of patients with pleural effusion.

\section{Methodology}
\label{sec2}
This section is divided in three subsections. In the first one, \ref{sec2.1}, we present a result that allows us to transform the problem discussed in (\ref{eq:H0X}) into an equivalent one, easier to handle, by using projections to reduce the multidimensional role of the covariate to a unidimensional one.

In subsection \ref{sec2.2} we show a methodology to test the equality of conditional ROC curves on a unidimensional problem  \citep[based on the one proposed in][]{Fanjul-Hevia2019}. Finally, in \ref{sec2.3}, we combine that methodology with the result obtained in \ref{sec2.1} to solve our original problem with multidimensional covariates. Both sections \ref{sec2.2} and \ref{sec2.3} include the statistic proposed to perform the test and a bootstrap algorithm to approximate its distribution.

\subsection{An equivalent problem}
\label{sec2.1}
In order to present the transformation of the problem, first we need to introduce the definition of \textit{the ROC curve conditioned to a pair} $({x^F}, {x^G}) \in R_{{X^F}}\times R_{{X^G}}$:
\begin{eqnarray}
ROC^{{x^F},{x^G}} (p) = 1- F(G^{-1}(1-p|{x^G})|{x^F}), \; p \in (0,1).
\label{eq:defROCxFxG}
\end{eqnarray}
This concept is very similar to the conditional ROC curve (\ref{eq:defROCx}): the only difference is that this new definition allows us to condition on different values for the diseased and healthy populations. In this case $x^F$ and $x^G$ are unidimensional, but the definition could be applied on a multidimensional case. Even if the interpretability of this new ROC curve is not very clear in practice, theoretically it does not present any problems (as it will not do its estimation), as the population of healthy and diseased are always considered to be independent.

The following result is the base for developing the test for comparing ROC curves with multidimensional covariates. 
It borrows the ideas in \cite{Escanciano2006} of using projections for reducing the dimension of the covariate in a regression context. Since here we are dealing with ROC curves, the dimension reduction is less straightforward and some adjustments are required, as each ROC depends on two cumulative distribution functions. To the best of our knowledge, the idea of using projections has not been considered in the context of ROC curves.


 Given $\bm{x}, \bm{\beta} \in \mathbb{R}^d$, $\bm{x'\beta}$ denotes the scalar product of the vectors $\bm{x}$ and $\bm{\beta}$. For now on, all the vectors representing the projections will be considered to be contained in the $d-$dimensional unit sphere $\mathbb{S}^{d-1} = \{\bm{\beta} \in \mathbb{R}^d : ||\bm{\beta}|| = 1\}$. This way we ensure that all possible directions are equally important.

\begin{lemma}
\label{lemma0} Assume  $\mathbb{E}|Y^F_k|<\infty$ and $\mathbb{E}|Y^G_k|<\infty$ for every $k \in\{1,\ldots,K\}$. 
Then, given a certain $\bm{x} \in \bm{R_X}$, and assuming dependence among the ROC curves (meaning the covariate is common for all the $K$ curves considered), then
\[ROC_1^{\bm{x}} (p)= \dots = ROC_K^{\bm{x}}(p) \; \text{ for all } p \in (0,1)  \; a.s.\]
if and only if
\[ROC_1^{(\bm{\beta}^F)' \bm{x},(\bm{\beta}^G)' \bm{x}} 
(p) = \dots = ROC_K^{(\bm{\beta}^F)' \bm{x},(\bm{\beta}^G)' \bm{x}} (p) \; \text{ for all } p \in (0,1) \; a.s. \; \text{ for any } \bm{\beta}^F , \bm{\beta}^G,\]
where $\bm{\beta}^F$ and $\bm{\beta}^G$ are $d-$dimensional coordinates in $\mathbb{S}^{d-1}$ that represent the directions of the projections. 
\end{lemma}

The proof of this Lemma can be found in the Appendix. Note that ${(\bm{\beta}^F)}' \bm{x}$ and ${\left(\bm{\beta}^G\right)}' \bm{x}$ are one-dimensional values. By using these ROC curves conditioned to a pair of projected covariates (as defined in \ref{eq:defROCxFxG}), the  problem is reduced to a one-dimensional covariate conditional ROC curve comparison test for each possible direction $\bm{\beta}^F$ and $\bm{\beta}^G$.

Thus, taking advantage of the result in Lemma~\ref{lemma0}, instead of testing for the null hypothesis (\ref{eq:H0X}), we may use this equivalent formulation to develop a methodology that, given a certain $\bm{x} \in R_{\bm{X}}$, tests
\begin{eqnarray}
H_0: ROC_1^{(\bm{\beta}^F)' \bm{x},(\bm{\beta}^G)' \bm{x}} (p) = \dots = ROC_K^{(\bm{\beta}^F)' \bm{x},(\bm{\beta}^G)' \bm{x}} (p) \; \text{ for all } p \in (0,1)  \; \forall \bm{\beta}^F, \bm{\beta}^G
\label{eq:H0forallproj}
\end{eqnarray}
against the general alternative $H_1:$ $H_0$ is not true. 
The notation $\forall$ will be used instead of `for any' to shorten the expression (this applies mainly in the proofs found in the Appendix).

In a first step, a statistic for testing the equivalence of these ROC curves is presented for a certain pair of fixed projections, and then that statistic is adapted to include all possible directions.

\subsection{Test for a one-dimensional covariate}
\label{sec2.2}
The objective in this section is to develop a test for the equivalent problem presented in Lemma~\ref{lemma0} for a fixed pair of projections $\bm{\beta}^F$ and $\bm{\beta}^G$.
Here  a test is presented for comparing two or more dependent ROC curves conditioned to two one-dimensional values.  Given the pair $(x^F, x^G) \in R_{X^F}\times R_{X^G}$, the aim is then to test
\begin{eqnarray}
H_0: ROC_1^{x^F, x^G} (p) = \cdots = ROC_K^{x^F, x^G} (p) \; \text{ for all } p \in (0,1)
\label{eq:H0xFxG}
\end{eqnarray}
against the general alternative $H_1: H_0$ is not true.

The samples available in this context are: 

\begin{itemize}
\item[-] $\{(X_{i}^F,Y_{1,i}^F,\dots, Y_{K,i}^F)\}_{i=1}^{n^F}$ an i.i.d. sample from the distribution of $(X^F,Y_1^F,\dots, Y_K^F)$,

\item[-] $\{(X_{i}^G,Y_{1,i}^G,\dots, Y_{K,i}^G)\}_{i=1}^{n^G}$ an i.i.d. sample from the distribution of $(X^G,Y_1^G, \dots, Y_K^G)$, 
\end{itemize}
with $n^F$ and $n^G$ the sample sizes of the diseased and healthy populations, respectively. Define 
$n=n^F+n^G$ as the total sample size used for the estimation of each conditional ROC curve (that will be the same for all $k \in \{1,\dots,K\}$). 
Note that both $X^F$ and $X^G$ are here one-dimensional covariates.

The method used for the estimation of the conditional ROC curves is based on the one proposed in \cite{Gonzalez-Manteiga2011a}, which relies on non-parametric location-scale regression models. To be more precise, for each $k = 1,\dots,K$, assume that
\begin{eqnarray}
Y_k^F = \mu_k^F(X^F) + \sigma_k^F(X^F)\varepsilon_k^F \label{eq:locscale1}\\
Y_k^G = \mu_k^G(X^G) + \sigma_k^G(X^G)\varepsilon_k^G 
\label{eq:locscale2}
\end{eqnarray}
where, for $D\in \{F,G\}$, $\mu_k^D(\cdot) = E(Y_k^D|X^D=\cdot)$ and $(\sigma_k^D)^2(\cdot) = Var(Y_k^D | X^D=\cdot)$ are the conditional mean and the conditional variance functions (both of them unknown smooth functions), and the error $\varepsilon_k^D$ is independent of $X^D$. 
 The dependence structure between the $K$ diagnostic variables is modelled by introducing a dependence structure between the errors: $(\varepsilon_1^D, \dots, \varepsilon_K^D)$ will follow a multivariate distribution function 
  with zero mean and a covariance matrix with ones in the diagonal.

Given this location-scale regression model structure for the diagnostic variables, the $k-$th ROC curve conditioned to a pair of values $({x^F}, {x^G}) \in R_{{X^F}}\times R_{{X^G}}$ can be expressed in terms of the marginal cumulative distribution functions of the errors, $H_k^F$ and $H_k^G$:
\begin{eqnarray}
ROC_k^{{x^F}, {x^G}}(p) = 1-H_k^F\left(\left(H_k^G\right)^{-1}(1-p)b_k(x^F,x^G)-a_k(x^F,x^G)\right),
\end{eqnarray}
where 
\[a_k(x^F,x^G) = \frac{\mu_k^F(x^F)- \mu_k^G(x^G)}{\sigma_k^F(x^F)} \quad \text{ and } \quad b_k(x^F,x^G) = \frac{\sigma_k^G(x^G)}{\sigma_k^F(x^F)}.\]
Thus, this $k-th$ conditional ROC curve can be estimated by
\begin{eqnarray}
{ \widehat{ROC}_k^{x^F,x^G}}(p) &=& 1- {\int}\hat{H}_k^F\left(\left(\hat{H}_k^G\right)^{-1}(1- p{+{ h_k} u }) \hat{b}_k({x^F,x^G}) -\hat{a}_k({x^F,x^G})\right) { \kappa(u) du},
\label{eq:EstROCxFxG}
\end{eqnarray}
where, for $D\in \{F,G\}$,
\begin{itemize}
\item  $\hat{H}_k^D(y) = (n^D)^{-1} \sum_{i=1}^{n^D} I(\hat{\varepsilon}_{k,i}^{D} \leq y)$, 
\item $\hat{\varepsilon}_{k,i}^{D} = \dfrac{Y_{k,i}^{D} -\hat{\mu}_k^D(X_i^{D})}{\hat{\sigma}_k^D (X_i^{D})}$,  with $i\in\{1,\cdots,n^D\}$,
\item $\hat{\mu}_k^D(x) = \sum_{i=1}^{n^D}  W_{k,i}^{D} (x, g_k^D) Y_{k,i}^{D}$  is a non-parametric estimator of $\mu_k^D(x)$  based on local weights $W_{k,i}^{D}(x,g_k^D)$  depending on a bandwidth parameter $g_k^D$,
\item $(\hat{\sigma}_k^D)^2(x) = \sum_{i=1}^{n^D}  W_{k,i}^{D} (x, g_k^D) [Y_{k,i}^{D}-\hat{\mu}_k^D(X_i^{D})]^2$  is a non-parametric estimator of  $(\sigma_k^D)^2(x)$. For simplicity we take the same bandwidth parameter $g_k^D$ that is used for the estimation of the regression function $\hat{\mu}_k^D(x)$,
\item  $W_{k,i}^{D} (x,g_k^D) = \dfrac{\kappa_{g_k^D}(x-X_{i}^D)}{\sum_{l=1}^{n^D} \kappa_{g_k^D}(x-X_{l}^D)}$   are Nadaraya-Watson-type weights, where $\kappa_{g_k^D}(\cdot) = \kappa(\cdot / g_k^D) /g_k^D$ and $\kappa$ is a probability density function symmetric around zero.
\item $\hat{a}_k (x^F,x^G)= \left(\hat{\mu}_k^F(x^F)-\hat{\mu}_k^G(x^G)\right) /\hat{\sigma}_k^F(x^F) $ and $\hat{b}_k(x^F,x^G) = \hat{\sigma}_k^G(x^G) / \hat{\sigma}_k^F(x^F)$.
\item $h_k$ is a bandwidth parameter responsible for the smoothness of the estimator. Its value does not seem to have a significant effect on the conditional ROC curve estimation.
\end{itemize}
This way of estimating the conditional ROC curve is similar to the one proposed in \cite{Gonzalez-Manteiga2011a}, with the difference that they condition the ROC curve on a single value $x$ and here we have a pair of values $x^F$ and $x^G$, each one of them related to the diseased and the healthy population, respectively. As both populations are independent, the adaptation of the methodology of \cite{Gonzalez-Manteiga2011a} to this case is straightforward.
 
Once we know how to estimate this doubly conditional ROC curve we can propose a test statistic for the test (\ref{eq:H0xFxG}):
\begin{eqnarray}
S^{x} = \sum_{k=1}^K \psi \left( \sqrt{n g_k} \{ \widehat{ROC_k}^{x^F,x^G}(p) - \widehat{ROC}_\bullet^{x^F,x^G}(p)\}\right),
\label{eq:SNx}
\end{eqnarray}
where:
\begin{itemize}

\item for $k\in \{1,\dots,K\}$, $g_k= \frac{n^F g_k^F + n^G g_k^G}{n}$, where $g_k^F$ and $g_k^G$ are bandwidth parameters involved in the estimation of the $k$-th conditional ROC curve.

\item for $k\in \{1,\dots,K\}$,
$ \widehat{ROC}_k^{x^F,x^G}(p)$ is the estimated conditional ROC curve given $(x^F,x^G)$, as seen in (\ref{eq:EstROCxFxG}),

\item $\widehat{ROC}_\bullet^{x^F,x^G} (p) = \left(\sum_{k=1}^K g_k  \right)^{-1}\sum_{k=1}^K  g_k  \widehat{ROC_k}^{x^F,x^G} (p)$ is a sort of weighted average of the $K$ conditional ROC curves.

\item $\psi$ is a real-valued function that measures the difference between each estimated conditional ROC curve and the weighted average of all of them. This function  may be similar to the ones used for the comparison of cumulative distribution functions (after all, a ROC curve can be viewed as a cumulative distribution function). For example, if one considers the $L_2$-measure, then the resulting test statistic is
 \[S_{L2}^{x} = \sum_{k=1}^K { n g_k} \int \left(   \widehat{ROC_k}^{x^F,x^G}(p) - \widehat{ROC}_\bullet^{x^F,x^G}(p)\right)^2 dp.\]
 On the other hand, when using the Kolmogorov-Smirnov criteria the resulting test statistic is
 \[S_{KS}^{x} = \sum_{k=1}^K \sqrt{ n g_k } \sup_{p} \left\vert  \widehat{ROC_k}^{x^F,x^G}(p) - \widehat{ROC}_\bullet^{x^F,x^G}(p)\right\vert.\]

\end{itemize}
The null hypothesis will be rejected for large values of $S^{x}$. In order to obtain the distribution of this statistic, a bootstrap algorithm is proposed. This bootstrap algorithm is adapted from the procedure proposed in \cite{Martinez-Camblor2012} and has been already used by \cite{Martinez-Camblor2013a} and by \cite{Fanjul-Hevia2019} in the context of ROC curves. The key of this algorithm is that 
\begin{eqnarray*}
T^{x} &=&  \sum_{k=1}^K \psi \left(   \sqrt{n g_k} \left\{ \left(\widehat{ROC}_k^{x^F,x^G}(p)- \widehat{ROC}_\bullet^{x^F,x^G}(p)\right) - \left(ROC_k^{x^F,x^G}(p)- ROC_\bullet^{x^F,x^G}(p)\right)\right\}\right),
\end{eqnarray*}
coincides with the statistic $S^{x}$ as long as the null hypothesis holds, 
where 
\[ROC_\bullet^{x^F,x^G}(p)=  \left(\sum_{k=1}^K  g_k \right)^{-1} \sum_{k=1}^K  g_k  ROC_k^{x^F,x^G}(p), \; 0<p<1.\] 
The quantity $T^{x}$ can be rewritten as
\begin{eqnarray}
T^{x} &=& \sum_{k=1}^K \psi \left(  \sum_{j=1}^K \sqrt{n g_j } \alpha_{kj} \{ \widehat{ROC}_j^{x^F,x^G}(p) - ROC_j^{x^F,x^G}(p)\}\right),
\label{eq:TNx}
\end{eqnarray}
where $\alpha_{kj} = I(k=j) - \sqrt{ g_k } \sqrt{g_j } \left(\sum_{i=1}^K  g_i \right)^{-1}$.  Note that, in general, $T^{x}$ cannot be computed from the data, as it depends on the unknown theoretical conditional ROC curves, but it is useful when applying the bootstrap algorithm.

The bootstrap algorithm suggested to approximate a p-value for this test is the following:

\begin{enumerate}
\item[A.1]  From the original samples, $\{(X_{i}^F,Y_{1,i}^F,\dots,Y_{K,i}^F)\}_{i=1}^{n^F}$ and $\{(X_{i}^G,Y_{1,i}^G,\dots,Y_{K,i}^G)\}_{i=1}^{n^G}$, compute the test statistic value (\ref{eq:SNx}), that we will denote by $s^x$.
\item[A.2] \label{Step2} 
For $b=1,\dots,B$, generate the bootstrap samples $\{(X_{i}^F,Y_{1,i}^{F,b*},\dots,Y_{K,i}^{F,b*})\}_{i=1}^{n^F}$ and\\ $\{(X_{i}^G,Y_{1,i}^{G,b*},\dots,Y_{K,i}^{G,b*})\}_{i=1}^{n^G}$ as follows:
\begin{enumerate}
\item[(i)] For each $D\in\{F,G\}$ , let $\left\{\left(\varepsilon_{1,i}^{D,b*},\dots,\varepsilon_{K,i}^{D,b*}\right)\right\}_{i = 1}^{n^D} $ be an i.i.d. sample from the { empirical cumulative multivariate distribution function of the original residuals}.

\item[(ii)] Reconstruct the bootstrap samples 
 $\{(X_{i}^D,Y_{1,i}^{D,b*},\dots,Y_{K,i}^{
D,b*})\}_{i=1}^{n^D}$ for each $D\in \{F,G\}$, where $Y_{k,i}^{D,b*} = \hat{\mu}_k^D(X_{k,i}^D) + \hat{\sigma}_k^D(X_{k,i}^D) \varepsilon_{k,i}^{D,b*}$.
\end{enumerate}

\item[A.3] \label{Step3}
Compute the test statistic based on the bootstrap samples, for $b=1,\dots,B$ using (\ref{eq:TNx}) as
\vskip-6pt
\begin{eqnarray*}
t^{x,b*} 
&=& \sum_{k=1}^K \psi \left(  \sum_{j=1}^K \sqrt{n g_j } \alpha_{kj} \{ \widehat{ROC}_j^{{x^F,x^G},b*}(p) - \widehat{ROC}_j^{x^F,x^G}(p)\}\right), \notag\\
\end{eqnarray*}
where $\widehat{ROC}_j^{x^F,x^G,b*}$ is the estimated $j-$th conditional ROC curve of the $b-$th bootstrap sample.
\item[A.4] The distribution of $S^x$ under the null hypothesis  (and thus, the distribution of $T^x$) is approximated by the empirical distribution of the values $\{t^{x,1*},\dots,t^{x,B*}\}$ and the p-value is approximated by
\[p-value = \frac{1}{B} \sum_{b=1}^B I(s^x \leq t^{x,b*}).\]
\end{enumerate}

In contrast with the usual bootstrap algorithms in testing setups, in this case the null hypothesis is not employed when  generating of the bootstrap samples (Step A.2), because replicating the null hypothesis of equal ROC curves is not a straightforward problem.
Instead, it is used in the computation of the bootstrap statistic (Step A.3) by using $T^x$ instead of $S^x$, that are equal under the null hypothesis. This particularity  also appears in the bootstrap algorithm of the next section.

There are two kind of bandwidth parameters that appear in the estimation of the $k-$th conditional ROC curve (\ref{eq:EstROCxFxG}), with $k\in \{1,\dots K\}$. The first one, $h_k$, is taken as $1/\sqrt{n}$, and the second ones, $g_k^F$ and $g_k^G$, are selected by least-squares cross-validation. Note that, for each bootstrap iteration, the bandwidth parameters could change, as their selection depends on the sample. However, $h_k$ remains constant, as we are choosing it in terms of the sample size, and that is the same for each bootstrap 
iteration. As for $g_k^F$ and $g_k^G$, for computational issues we have decided to compute them on step A.1 using the original sample, and then apply the same bandwidths for all the bootstrap estimations. The cross-validation method can be very time-consuming, and this simplification prevents the simulations to become infeasible.

\subsection{Test for a multi-dimensional covariate}
\label{sec2.3}


Once having seen a strategy for testing (\ref{eq:H0forallproj}) for only one pair of fixed directions, the idea now is to modify the previous procedure so the new statistic  takes into account all the possible directions that $\bm{\beta}^F$ and $\bm{\beta}^G$ can take. For that purpose, consider the test statistic
\begin{eqnarray}
D_S^{\bm{x}} = \int_{\mathbb{S}^{d-1}} \int_{\mathbb{S}^{d-1}}   S^{(\bm{\beta}^F)' \bm{x},(\bm{\beta}^G)'\bm{x}} d\bm{\beta}^F d \bm{\beta}^G,
\label{eq:DN1}
\end{eqnarray}
where $d\bm{\beta}^F$ and $d\bm{\beta}^G$ represent the uniform density on  the sphere  of dimension $d$, $\mathbb{S}^{d-1}$. This ensures that all directions are equally important. 

 The expression $S^{(\bm{\beta}^F)'\bm{x},(\bm{\beta}^G)'\bm{x}}$ is equal to the statistic used  in (\ref{eq:SNx}) for testing the equality of $K$ ROC curves when conditioned to the value of the pair $\left((\bm{\beta}^F)'\bm{x},(\bm{\beta}^G)'\bm{x}\right)$, that is, 
\[S^{(\bm{\beta}^F)'\bm{x},(\bm{\beta}^G)'\bm{x}} = \sum_{k=1}^K \psi \left( \sqrt{n g_k } \{ \widehat{ROC_k}^{(\bm{\beta}^F)'\bm{x},(\bm{\beta}^G)'\bm{x}}(p) - \widehat{ROC}_\bullet^{(\bm{\beta}^F)'\bm{x},(\bm{\beta}^G)'\bm{x}}(p)\}\right). \]
Note that, in this context with $d-$dimensional covariates, the samples  are $\{(\bm{X}_{i}^F,Y_{1,i}^F,\dots, Y_{K,i}^F)\}_{i=1}^{n^F}$ and  $\{(\bm{X}_{i}^G,Y_{1,i}^G,\dots, Y_{K,i}^G)\}_{i=1}^{n^G}$ , with $\bm{X}_{i}^F = (X_{1,i}^F, \cdots, X_{d,i}^F)'$ and $\bm{X}_{i}^G = (X_{1,i}^G, \cdots, X_{d,i}^G)'$.

In practice, as it is done in \cite{Colling2017}, to compute the test statistic $D_S^{\bm{x}}$ random directions $\bm{\beta}_1^F, \dots, \bm{\beta}_{n_\beta}^F$ and $\bm{\beta}_1^G, \dots, \bm{\beta}_{n_\beta}^G$ are drawn uniformly from $\mathbb{S}^{d-1}$, where $n_{\bm{\beta}}$ is the number of random directions considered (the same number of directions is taken for $\bm{\beta}^F$ and for $\bm{\beta}^G$). With them, the approximated  statistic is
\begin{eqnarray}
\tilde{D}_S^{\bm{x}} = \frac{1}{n_{\bm{\beta}}^2} \sum_{r=1}^{n_{\bm{\beta}}} \sum_{l=1}^{n_{\bm{\beta} }}  S^{(\bm{\beta}_r^F)'\bm{x},(\bm{\beta}_l^G)' \bm{x}}.
\label{ref:DNap}
\end{eqnarray}

In order to obtain the distribution of the statistic, a bootstrap algorithm (similar to the one described in the previous section) is proposed. To do so, the following expression is introduced:
\begin{eqnarray}
D_{T}^{\bm{x}} = \int_{\mathbb{S}^{d-1}} \int_{\mathbb{S}^{d-1}}   T^{(\bm{\beta}^F)' \bm{x},(\bm{\beta}^G)' \bm{x}} d\bm{\beta}^F d \bm{\beta}^G,
\label{eq:TN1}
\end{eqnarray}
where $T^{(\bm{\beta}^F)' \bm{x},(\bm{\beta}^G)' \bm{x}}$ is the same as in  (\ref{eq:TNx}), but for the conditioning values of $\left((\bm{\beta}^F)' \bm{x},(\bm{\beta}^G)' \bm{x}\right)$:
\begin{eqnarray*}
T^{(\bm{\beta}^F)' \bm{x},(\bm{\beta}^G)' \bm{x}} =  \sum_{k=1}^K \psi \left(  \sum_{j=1}^K \sqrt{n g_j} \alpha_{kj} \{ \widehat{ROC}_j^{(\bm{\beta}^F)' \bm{x},(\bm{\beta}^G)' \bm{x}}(p) - ROC_j^{(\bm{\beta}^F)' \bm{x},(\bm{\beta}^G)' \bm{x}}(p)\}\right).
\end{eqnarray*}
As it happened in (\ref{eq:TNx}), $T^{(\bm{\beta}^F)' \bm{x},(\bm{\beta}^G)' \bm{x}}$ cannot be computed without knowing the true distribution of the diagnostic markers. However, it can be computed in the bootstrap algorithm below, and there $D_{T}^{\bm{x}}$ is approximated by
\begin{eqnarray}
\tilde{D}_{T}^{\bm{x}} = \frac{1}{n_{\bm{\beta}}^2} \sum_{r=1}^{n_{\bm{\beta}}} \sum_{l=1}^{n_{\bm{\beta}}}   T^{(\bm{\beta}_r^F)' \bm{x},(\bm{\beta}_l^G)' \bm{x}}.
\label{ref:DTNap}
\end{eqnarray}

As happened before, for two given projections $\bm{\beta}^F$ and $\bm{\beta}^G$, $S^{(\bm{\beta}^F)'\bm{x},(\bm{\beta}^G)' \bm{x}}$ and $T^{(\bm{\beta}^F)'\bm{x},(\bm{\beta}^G)'\bm{x}}$ coincide as long as the null hypothesis holds, and thus the same happens with 
$D_S^{\bm{x}} $ and $D_{T}^{\bm{x}} $.

%

Taking into account these approximations, the resulting bootstrap algorithm goes as follows:

\begin{itemize}
\item[B.1] Draw $n_{\bm{\beta}}$ random directions $\bm{\beta}_1^F, \dots, \bm{\beta}_{n_\beta}^F$ and $\bm{\beta}_1^G, \dots, \bm{\beta}_{n_\beta}^G$ uniformly from $\mathbb{S}^{d-1}$.
\item[B.2] For each random directions $\bm{\beta}_r^F$ and $\bm{\beta}_l^G$ (with $r, {l}\in \{1, \dots,n_{\bm{\beta}}\}$) , consider the sample \\ $\left\{\left((\bm{\beta}_r^F)' \bm{X}_{i}^F,Y_{1,i}^F,\dots,Y_{K,i}^F\right)\right\}_{i=1}^{n^F}$ and $\left\{\left((\bm{\beta}_l^G)' \bm{X}_{i}^G,Y_{1,i}^G,\dots,Y_{K,i}^G\right)\right\}_{i=1}^{n^G}$  and the conditioning values\\ $\left( (\bm{\beta}_r^F)' \bm{x}, (\bm{\beta}_l^G)' \bm{x}\right)$. With them, following steps A.1--A.3 of the bootstrap algorithm of the previous subsection, compute the value of $s^{(\bm{\beta}_r^F)' \bm{x}, (\bm{\beta}_l^G)' \bm{x}}$ and the $B$ corresponding $t^{(\bm{\beta}_r^F)' \bm{x}, (\bm{\beta}_l^G)' \bm{x}, b*}$.
\item[B.3] Compute $\tilde{d}_S^{\bm{x}} = \frac{1}{n_{\bm{\beta}}^2} \sum_{r=1}^{n_\beta} \sum_{l=1}^{n_\beta}   s^{ (\bm{\beta}_r^F)' \bm{x},(\bm{\beta}_l^G)' \bm{x}}$ and $\tilde{d}_{T}^{\bm{x},b*} = \frac{1}{n_{\bm{\beta}}^2} \sum_{r=1}^{n_{\bm{\beta}}} \sum_{l=1}^{n_{\bm{\beta}}}   t^{(\bm{\beta}_r^F)'\bm{x},(\bm{\beta}_l^G)' \bm{x}, b*}$ as in (\ref{ref:DNap}) and (\ref{ref:DTNap}).
\item[B.4] Approximate the p-value of the test by:
\[p-value = \frac{1}{B} \sum_{b=1}^B I(\tilde{d}_S^{\bm{x}} \leq \tilde{d}_{T}^{\bm{x},b*}).\]
\end{itemize}

\begin{remark}
\label{remark1}
Note that $n_{\bm{\beta}}$ represents the number of random directions drawn from $\mathbb{S}^{d-1}$ considered for the approximation of (\ref{ref:DNap}) and (\ref{ref:DTNap}), but that, in fact, we are using $n_{\bm{\beta}}^2$ different combination of pairs $(\bm{\beta}^F, \bm{\beta}^G)\in \mathbb{S}^{d-1}\times\mathbb{S}^{d-1}$  to make that approximation. This could become a problem from the computational point of view, as the complexity of the problem increases very fast when increasing the value of $n_{\bm{\beta}}$. 

As an alternative, we could consider using
\begin{eqnarray*}
D_S^{\bm{x}} = \int_{\mathbb{S}^{d-1}\times\mathbb{S}^{d-1}}  S^{(\bm{\beta}^F)' \bm{x},(\bm{\beta}^G)'\bm{x}} d\bm{\beta}^F \bm{\beta}^G,
\end{eqnarray*}
instead of statistic  (\ref{eq:DN1}), 
where $d\bm{\beta}^F\bm{\beta}^G$ represents the uniform density on  the torus  of dimension $d$, $\mathbb{S}^{d-1}\times\mathbb{S}^{d-1}$. This ensures, as before, that all pairs of directions are equally important. Thus, in practice, instead of using the approximation (\ref{ref:DNap}) we could consider
\begin{eqnarray*}
\hat{D}_S^{\bm{x}} = \frac{1}{m_{\bm{\beta}}} \sum_{r=1}^{m_{\bm{\beta}}}  S^{(\bm{\beta}_r^F)'\bm{x},(\bm{\beta}_r^G)' \bm{x}},
\end{eqnarray*}
where $(\bm{\beta}_1^F,\bm{\beta}_1^G), \dots, (\bm{\beta}_{m_\beta}^F,\bm{\beta}_{m_\beta}^G)$  are pairs of random directions drawn uniformly from $\mathbb{S}^{d-1} \times \mathbb{S}^{d-1} $, and where 
$m_{\bm{\beta}}$ would represent here the same as $n_{\bm{\beta}}^2$ before, with the advantage that it allows for more flexibility because it can assume non-squared values. A similar adaptation could be applied for the approximation of $D_T^{\bm{x}}$ in (\ref{eq:TN1}).


\end{remark}

\begin{remark}
\label{remark2}
In the literature we can find papers, like for example  \cite{Cuesta-Albertos2007} or \cite{Cuesta-Albertos2019}, that use only one random projection. The main idea is to perform the test at hand for a randomly selected projection instead of for all possible projections. 
The use of projections results in a dimension reduction (as desired), and, despite being a procedure that may produce less powerful tests,  the use of one single projection results in a reduction of the computational cost.

Following that idea, instead of testing the equality of covariate-projected ROC curves for all possible projections,  we could  test the equality of covariate-projected ROC curves for some random pair of projections given a certain $\bm{x} \in R_{\bm{X}}$, meaning: 


\begin{eqnarray}
H_0: ROC_1^{(\bm{\beta}^F)' \bm{x},(\bm{\beta}^G)' \bm{x}} = \dots = ROC_K^{(\bm{\beta}^F)' \bm{x},(\bm{\beta}^G)' \bm{x}}  \; \text{ for some } \bm{\beta}^F, \bm{\beta}^G.
\label{eq:H0forrandproj}
\end{eqnarray}

The equivalence between this hypothesis and the one of interest in this paper given in (\ref{eq:H0X}) still needs theoretical justification. However, it is a possibility worth studying, if only for computational reasons
. A way of perform this approach could be to consider the proposed methodology for $n_{\bm{\beta}} =1$.
\end{remark}

\section{Simulations}
\label{sec3}

In order to analyse the performance of the proposed  methodology, simulations were run for the comparison of several dependent conditional ROC curves.  On a first stage, these simulations were focused on analysing the behaviour of the unidimensional test described in Section~\ref{sec2.2}, but we do not display them here, as they are very similar to the ones that can be found in \cite{Fanjul-Hevia2019}. Instead, we show the results for several scenarios (first under the null hypothesis and then under the alternative) in which we compare $K$ ROC curves (with $K\in\{2,3\}$) conditioned to a $d-$dimensional covariate (with $d\in\{2,3\}$). 

All the curves used in the simulation study were  drawn from location-scale regression models similar to the ones presented in (\ref{eq:locscale1}) and (\ref{eq:locscale2}), only that, in this case, the regression and the conditional standard deviation functions are for $d-$dimensional covariates. The construction of those curves is summarized in Table~\ref{table:SimulROCx}, were all the different conditional mean and conditional standard deviation functions are displayed.

\begin{table}[!t]
  \centering
\begin{tabular}{m{1.75cm} m{1.85cm} m{6cm} m{4.5cm}}
  \hline  
 \thead{\textbf{Covariate}} & \thead{\textbf{ROC curves}} & \thead{\textbf{Regression functions}}& \thead{\textbf{Conditional standard }\\\textbf{deviation functions}} \\
  \hline
 & \centering $ ROC_1^{\bm{x}}$
  	&
		$\mu_1^F (\bm{x})=  \sin(0.5  \pi  x_1) + 0.1x_2$ 
		
		$\mu_1^G (\bm{x})= 0.5x_1  x_2$
	&

		$\sigma_1^F (\bm{x})=  0.5 + 0.5x_1$ 
		
		$\sigma_1^G (\bm{x})= 0.5 + 0.5x_1$
 	\\
\centering $\bm{x} =\begin{pmatrix} x_1 \\x_2 \end{pmatrix}$ &\centering  $ ROC_2^{\bm{x}}$
  	&
		$\mu_2^F (\bm{x})=  0.3 + \sin(0.5  \pi  x_1) + 0.1 x_2$ 
		
		$\mu_2^G (\bm{x})= 0.5x_1  x_2$
	&

		$\sigma_2^F (\bm{x})=  0.5 + 0.5x_1$ 
		
		$\sigma_2^G (\bm{x})= 0.5 + 0.5x_1$
	\\
& \centering  $ ROC_3^{\bm{x}}$
  	&
		$\mu_3^F (\bm{x})=  \sin(0.5  \pi x_1) + 0.1 x_2$ 
		
		$\mu_3^G (\bm{x})= -0.3 + 0.4 x_2 + 0.5 x_1  x_2$
	&

		$\sigma_3^F (\bm{x})=  0.5 + 0.5x_1$ 
		
		$\sigma_3^G (\bm{x})= 0.5 + 0.5x_1$
 	\\  
  \hline
& \centering  $ ROC_4^{\bm{x}}$
  	&
		$\mu_4^F (\bm{x})=  \sin(0.5 \pi x_1) + 0.1 x_2 + 0.5  x_3,$ 
		
		$\mu_4^G (\bm{x})= 0.5 x_1  x_2 + x_3$
	&

		$\sigma_4^F (\bm{x})=  0.5 + 0.1 x_3
,$ 
		
		$\sigma_4^G (\bm{x})= 0.5 + 0.1 x_3$
 	\\
\centering  $\bm{x} =\begin{pmatrix} x_1 \\x_2\\x_3 \end{pmatrix}$
& \centering  $ ROC_5^{\bm{x}}$
  	&
		$\mu_5^F (\bm{x})=  \sin(0.5 \pi x_1) + 0.1 x_2 + 0.5  x_3,$
		
		$\mu_5^G (\bm{x})= x_1  x_2 + x_3$
	&

		$\sigma_5^F (\bm{x})= 0.5 + 0.1 x_3$

		$\sigma_5^G (\bm{x})= 0.5 + 0.1 x_3$
	\\
 &\centering  $ ROC_6^{\bm{x}}$
  	&
		$\mu_6^F (\bm{x})=  \sin(0.5 \pi x_1) + 0.1 x_2 + 0.5  x_3,$
		
		$\mu_6^G (\bm{x})= -0.3 + 0.5 x_1  x_2 + x_3$
	&

		$\sigma_6^F (\bm{x})= 0.5 + 0.2 x_2 + 0.3  x_3$

		$\sigma_6^G (\bm{x})= 0.5 + 0.1 x_3$
 	\\  
  \hline
\end{tabular}
\caption{Conditional mean and conditional standard deviation functions of the conditional ROC curves considered in the simulation study.}
\label{table:SimulROCx}
\end{table}
The regression errors 
 were considered to have multivariate normal distribution with zero mean, variance one and correlation $\rho$  for all the models. 

In all scenarios the covariates $X_1^F$, $X_1^G$, $X_2^F$, $X_2^G$, $X_3^F$ and $X_3^G$ are uniformly distributed in the unit interval. Thus, the value of the multidimensional covariate $\bm{x}$ at which the conditional ROC curves should be compared is contained in $[0,1]^d$. Particularly, the comparisons are made for $\bm{x} = (0.5,0.6)'$ and for $\bm{x} = (0.5,0.6,0.5)'$, for $d=2$ and $d=3$, respectively.  

The study contains simulations for different sample sizes $(n^F,n^G) \in \{(100,100)$, $(250,150)$, $(250,350)\}$ and different values of 
$\rho$ that represent different possible degrees of correlation between the diagnostic variables under comparison ($\rho \in \{-0.5, 0, 0.5\}$). 

Moreover, two different functions $\psi$ were considered for the construction of $S^{(\bm{\beta}^F)' \bm{x}, (\bm{\beta}^G)' \bm{x}}$: one based on the
$L_2-$measure and the other one based on the Kolmogorov-Smirnov criterion (from now on denoted by $L_2$ and $KS$ respectively). The number of iterations used in the bootstrap algorithm was 200, and 500
data sets were simulated  to compute the proportion of rejection in each scenario. 

Furthermore, the number of directions  that was used for approximating the test statistic $D_S^{\bm{x}}$
 was taken as $n_{\bm{\beta}}=5$ (as mentioned in Remark~\ref{remark1}, notice that this means that $n_{\bm{\beta}}^2=$25 different pairs of directions were considered). 

\subsection{Level of the test}
\label{sec3.1}

The scenarios that were considered for calibrating the level of the test (by comparing the same conditional ROC curves) are represented in Table~\ref{table:level}.

\begin{figure}[!h]
\centering\includegraphics[scale=0.41]{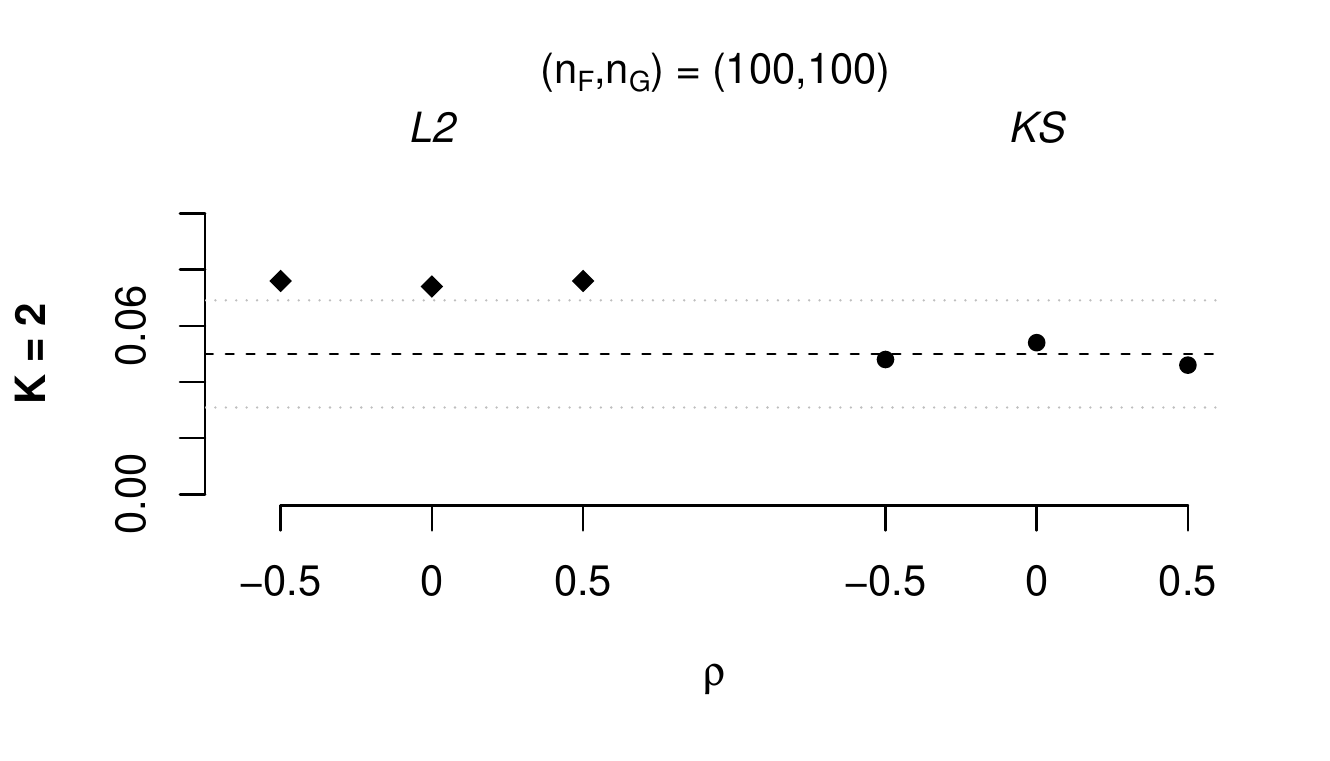}
\centering\includegraphics[scale=0.41]{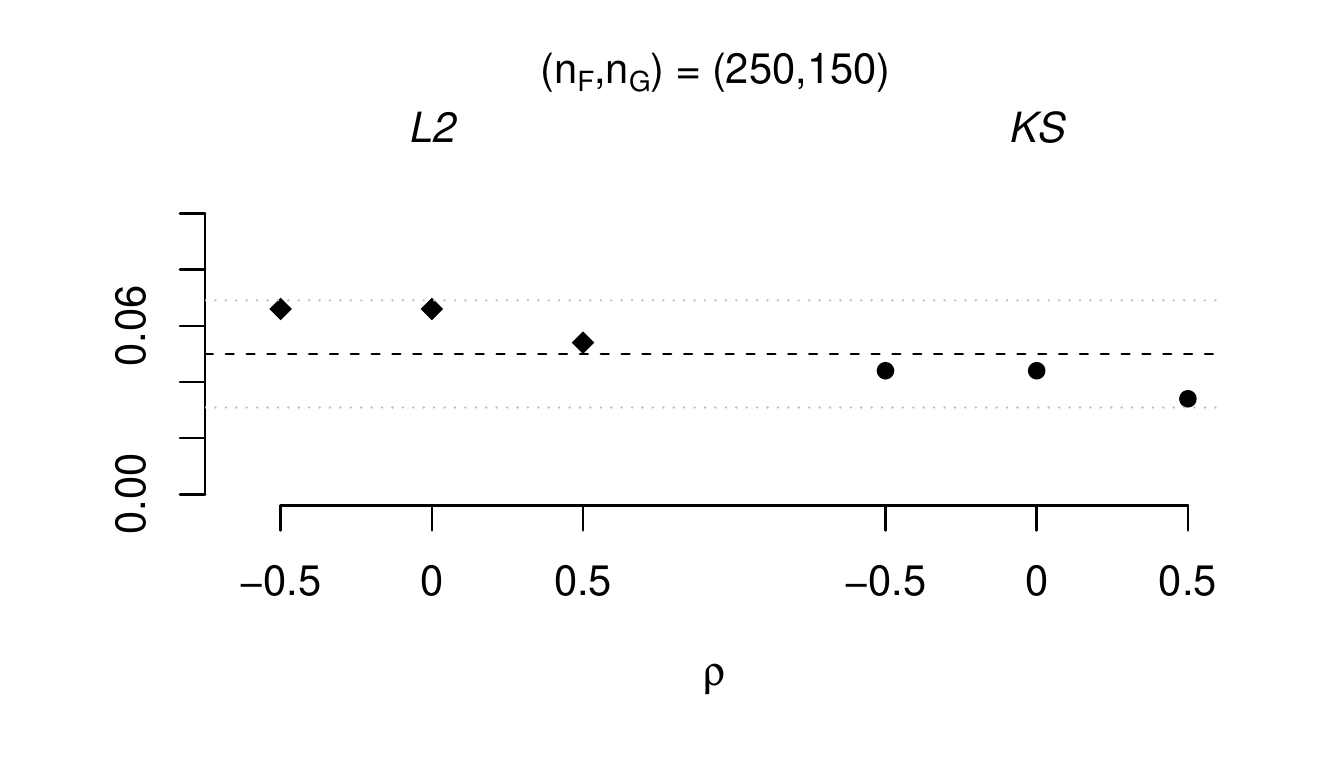}
\centering\includegraphics[scale=0.41]{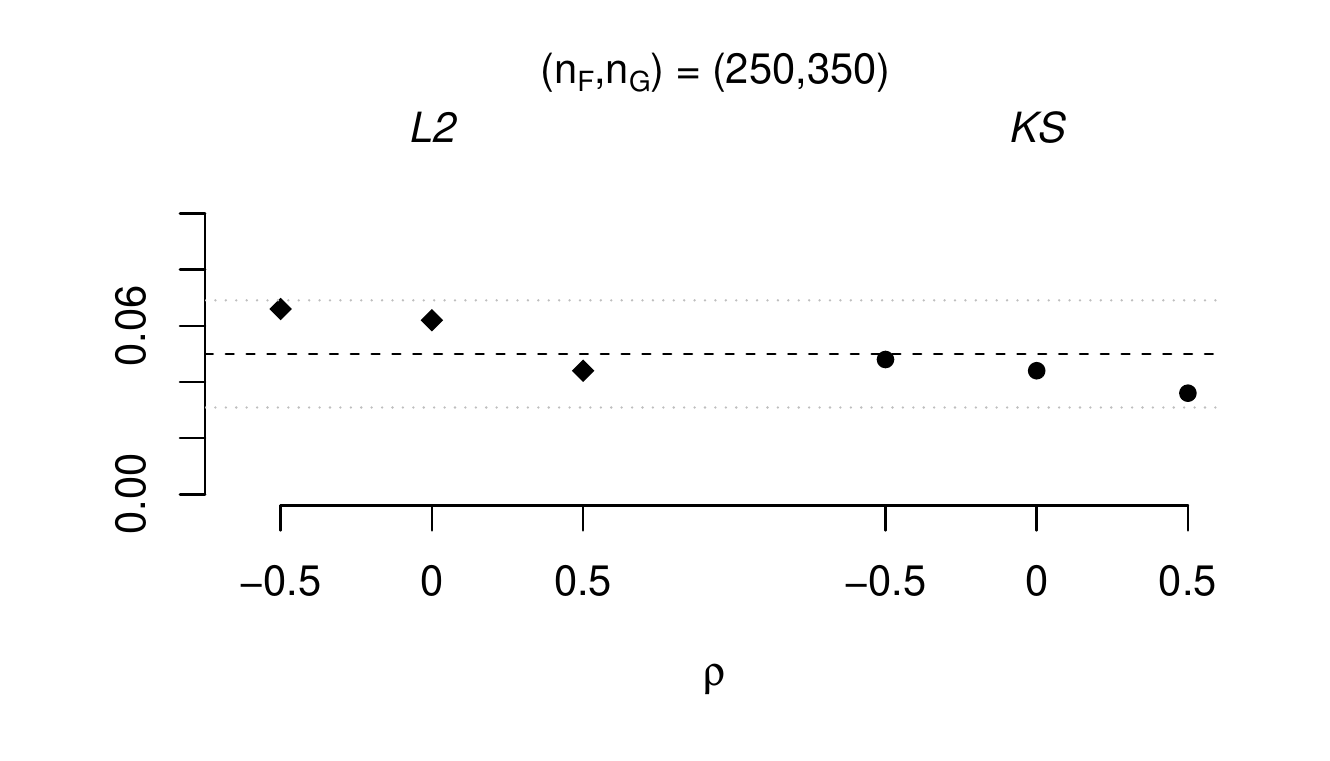}

\includegraphics[scale=0.41]{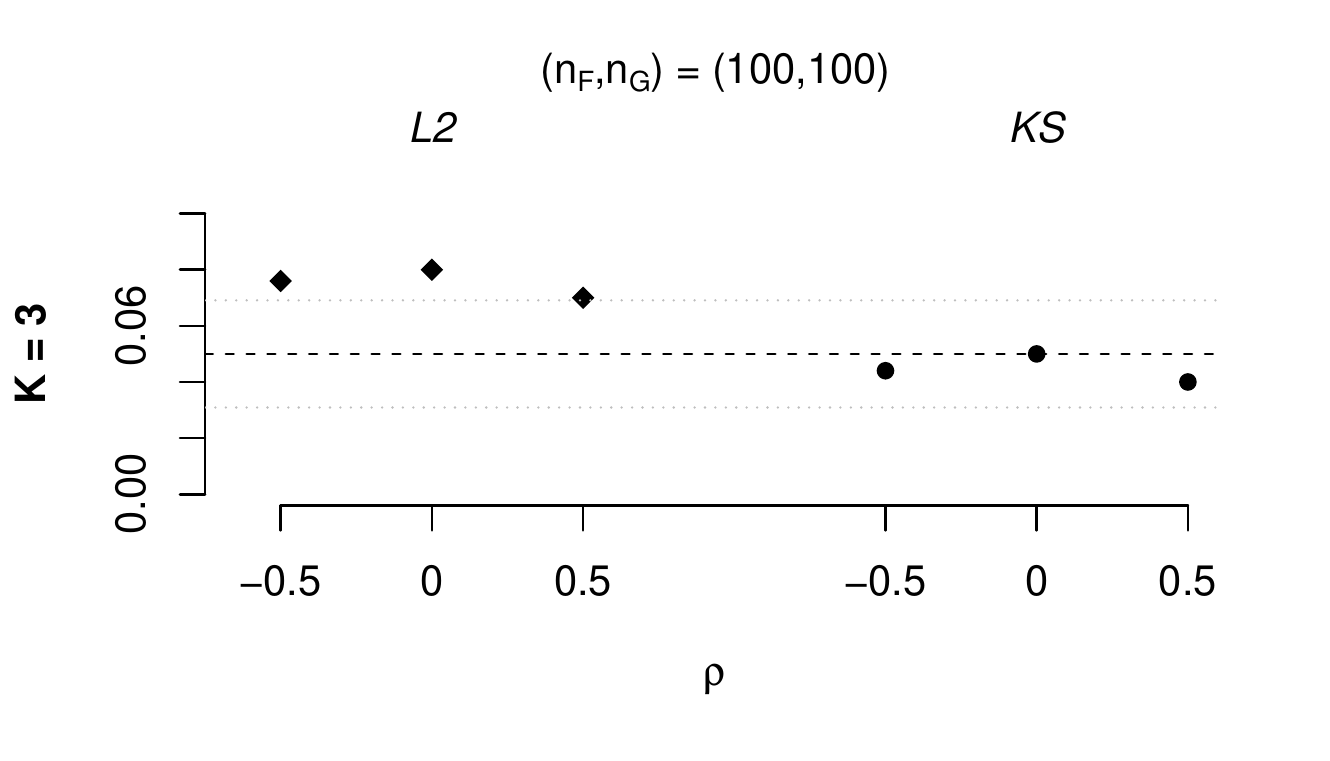}
\centering\includegraphics[scale=0.41]{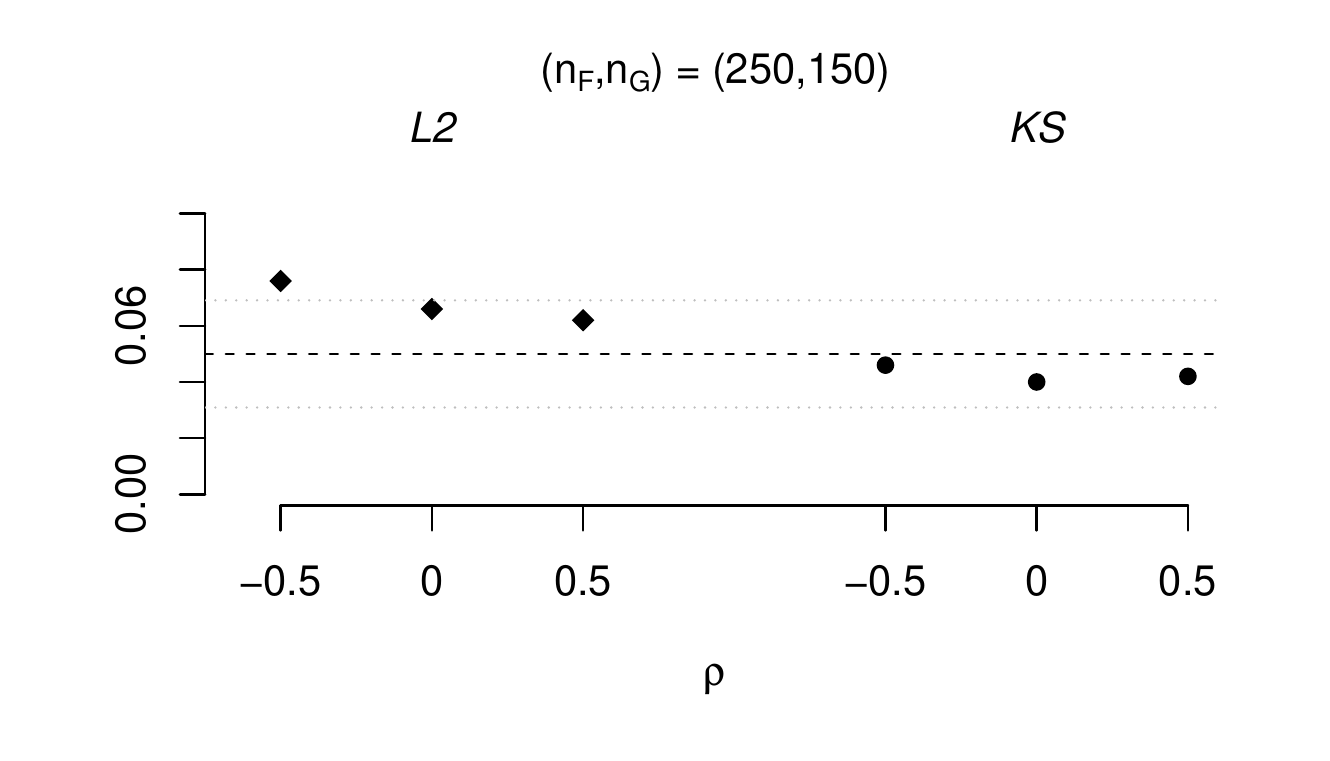}
\centering\includegraphics[scale=0.41]{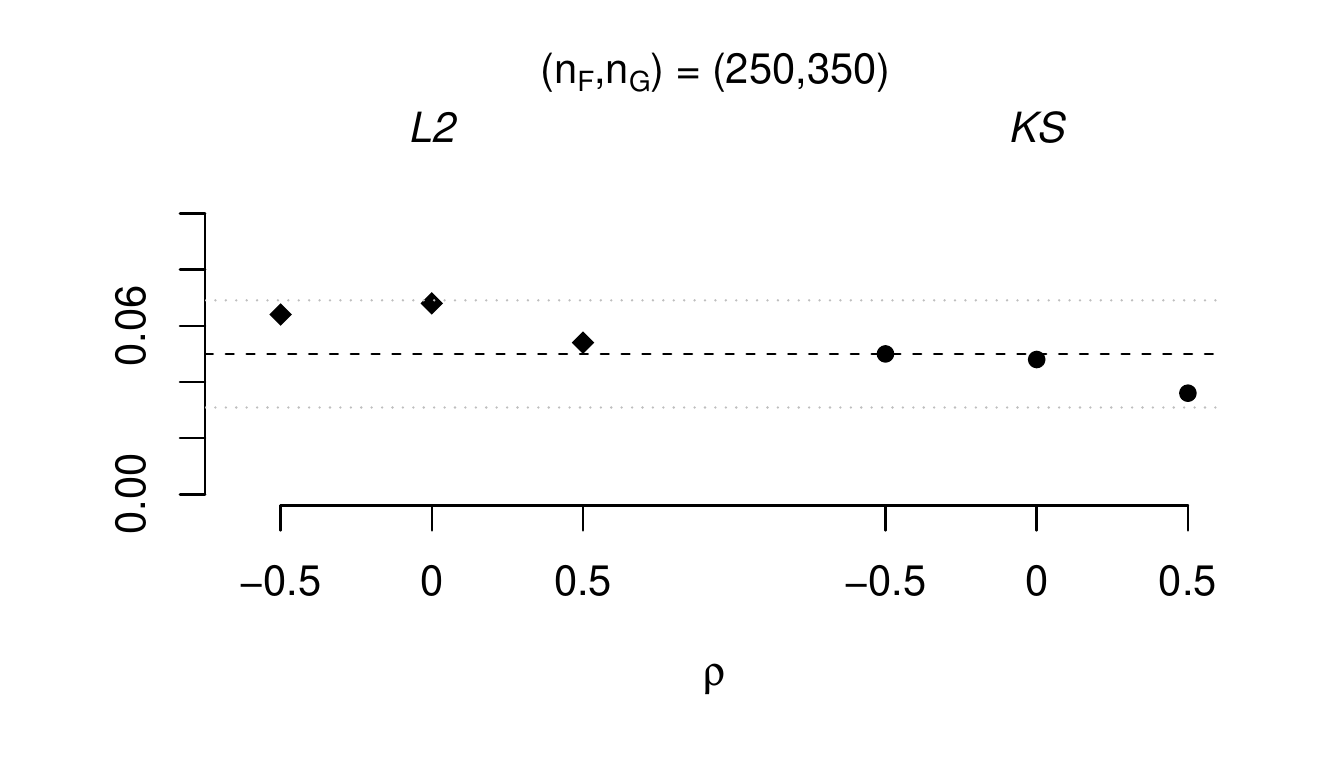}
\caption{Estimated proportion of rejection under the null hypothesis and the corresponding limits of the critical region (in gray) for the level 0.05 (dotted black line) with $d=2$ and $n_{\bm{\beta}} = 5$ for different sample sizes and different $\rho$.}
\label{fig:levelX2n5}
\end{figure}

\begin{figure}[!]
\centering\includegraphics[scale=0.41]{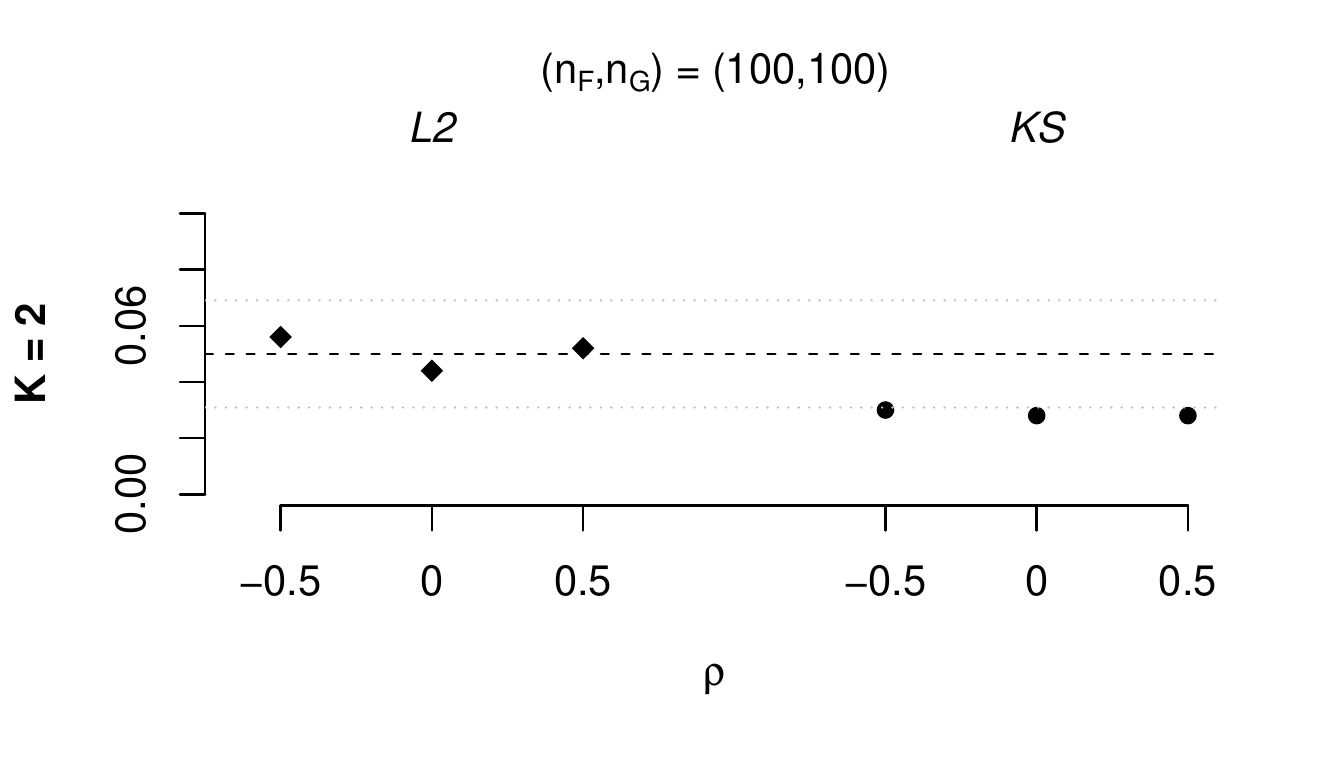}
\centering\includegraphics[scale=0.41]{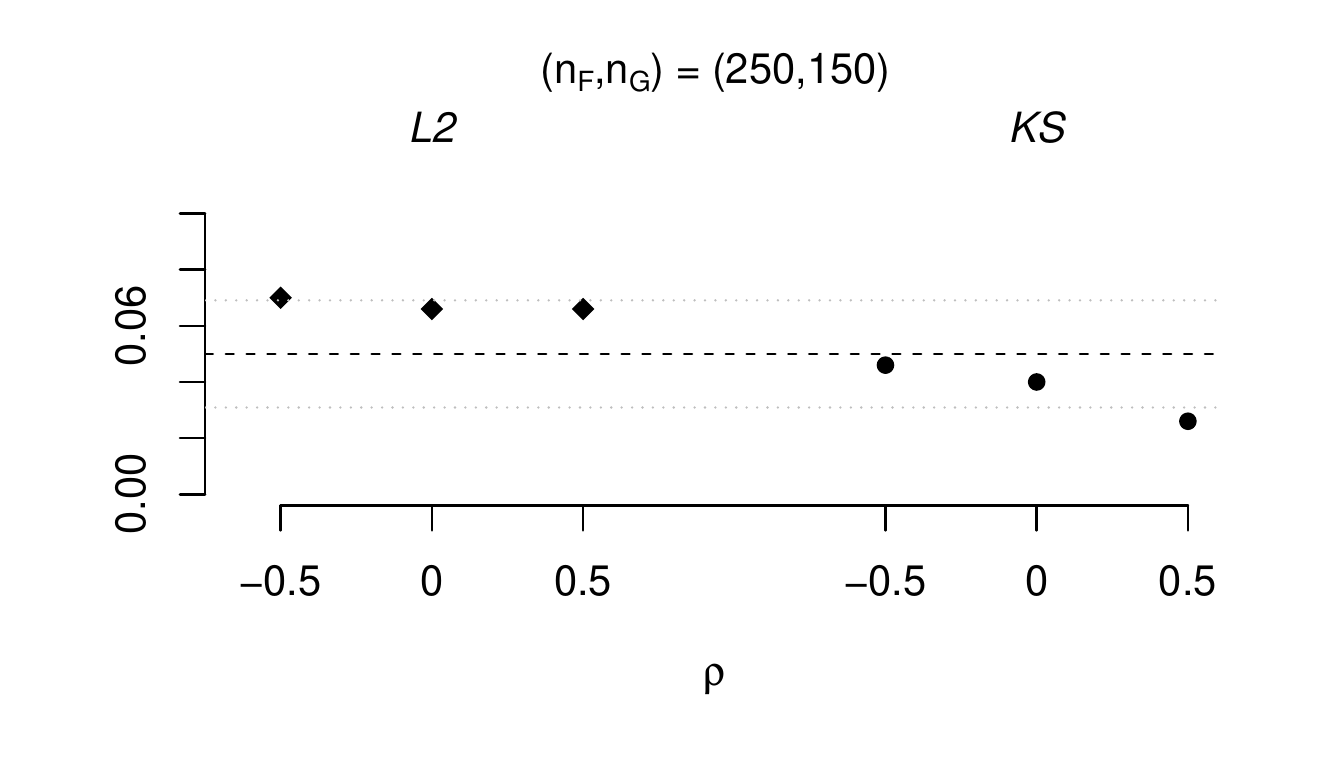}
\centering\includegraphics[scale=0.41]{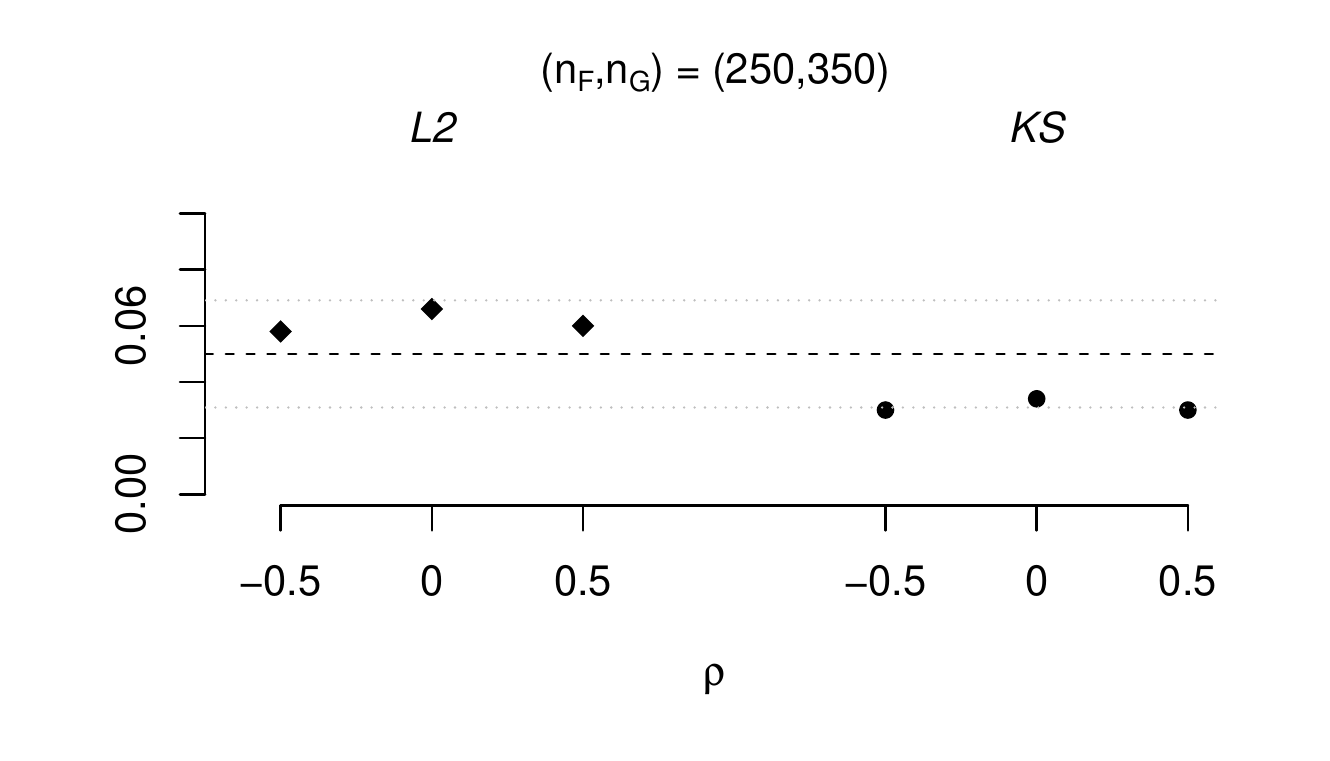}

\includegraphics[scale=0.41]{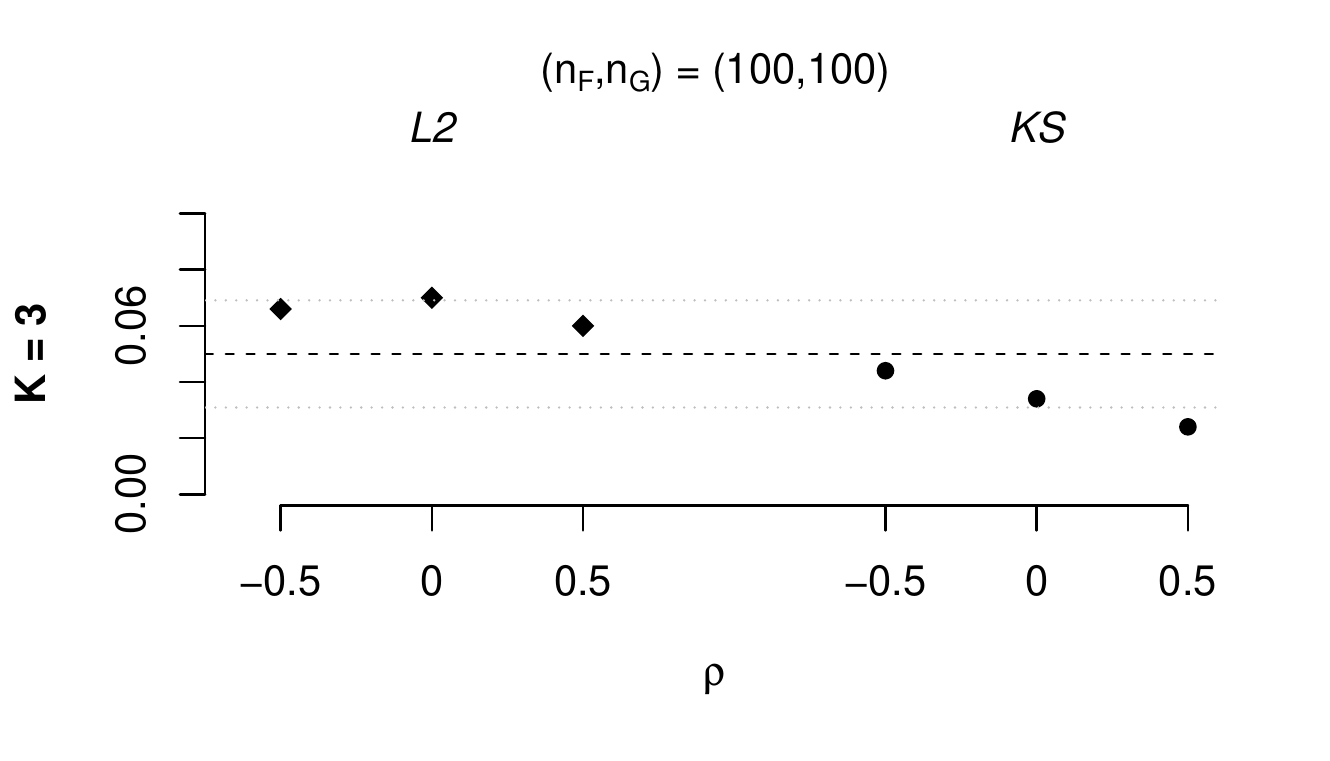}
\centering\includegraphics[scale=0.41]{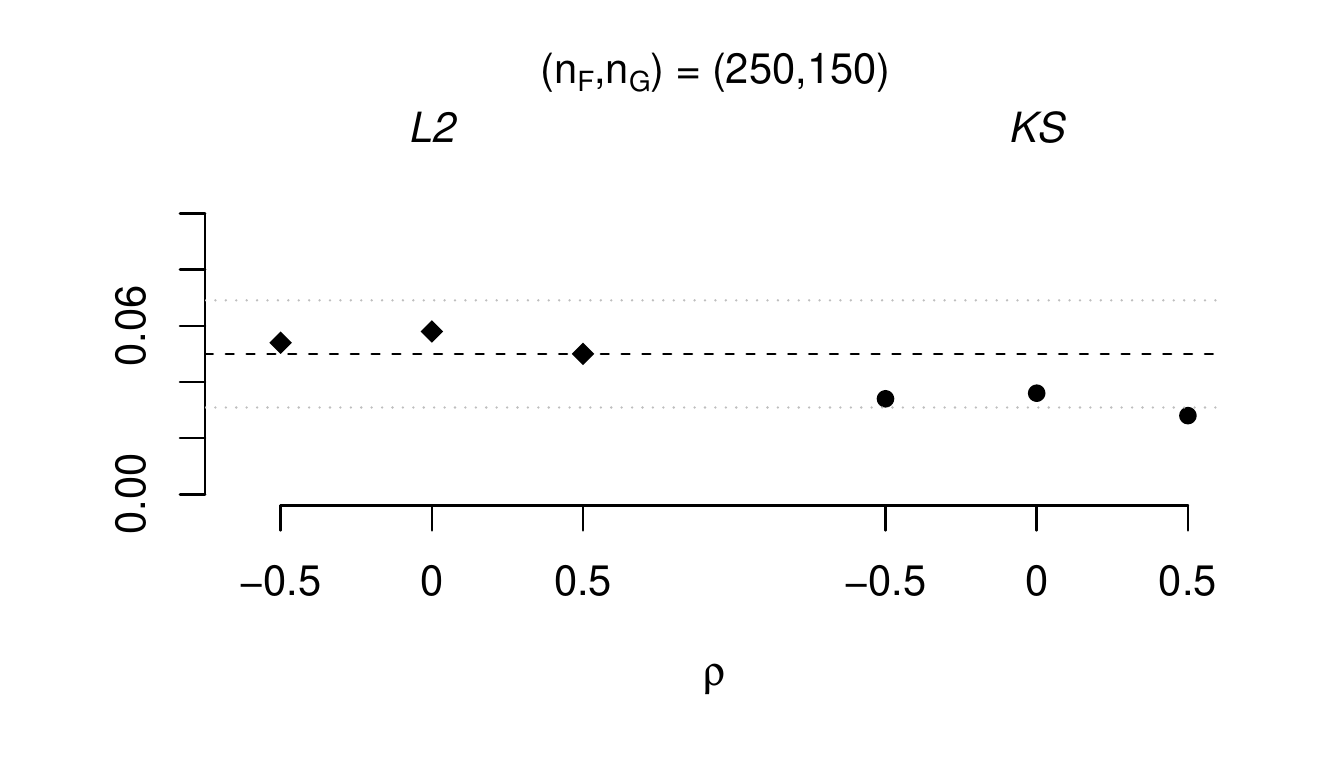}
\centering\includegraphics[scale=0.41]{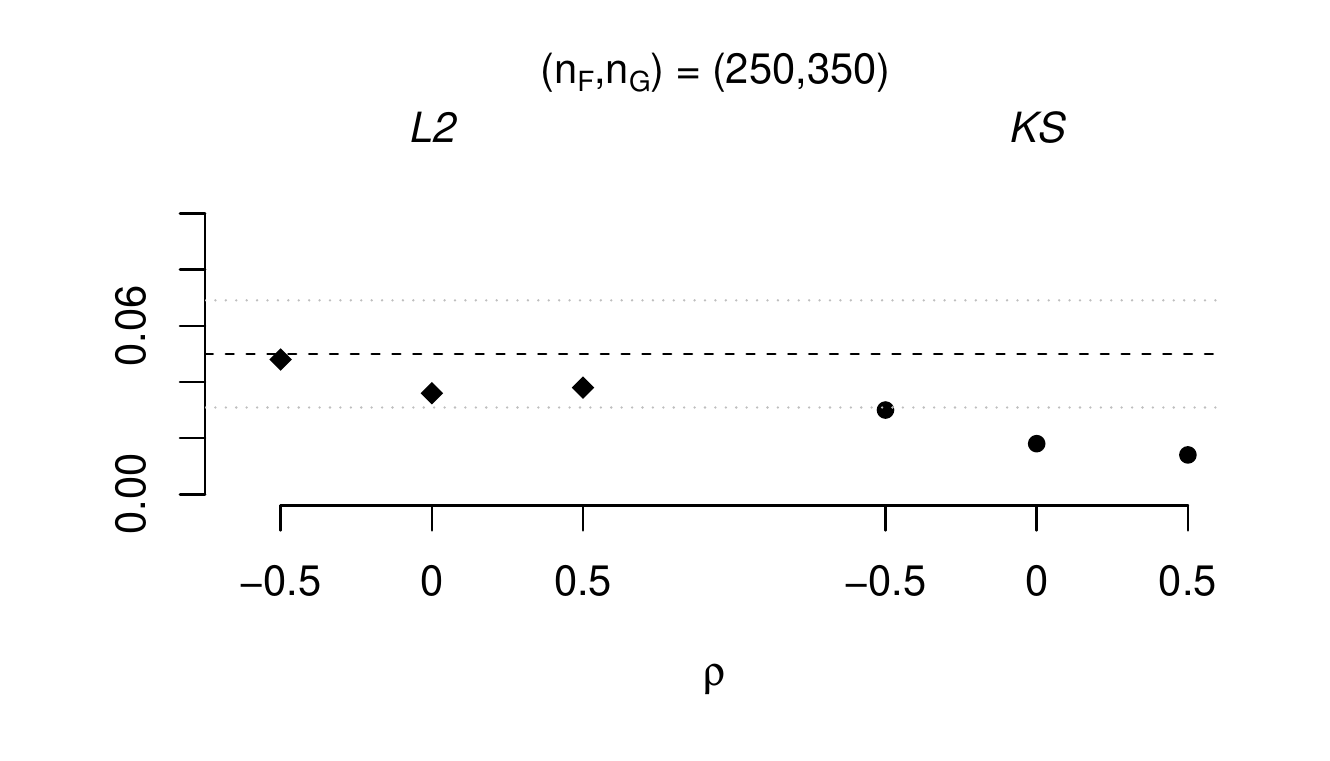}
\caption{Estimated proportion of rejection under the null hypothesis and the corresponding limits of the critical region (in gray) for the level 0.05 (dotted black line) with $d=3$ and $n_{\bm{\beta}} = 5$ for different sample sizes and different $\rho$.}
\label{fig:levelX3n5}
\end{figure}

\begin{table}[!t]
\begin{center}
\begin{tabular}{ m{1.5cm}  m{6cm} m{6.5cm}}
\hline
\centering   & \centering \textbf{$2-$dimensional covariate} &   \textbf{$3-$dimensional covariate}\\
\hline
\vspace{0.2cm}
\centering $K=2,3$ & \centering  $ROC_1^x$ &  \qquad \qquad \qquad $ROC_4^x$ \\
&
\vspace{0.1cm}
\includegraphics[scale=0.65]{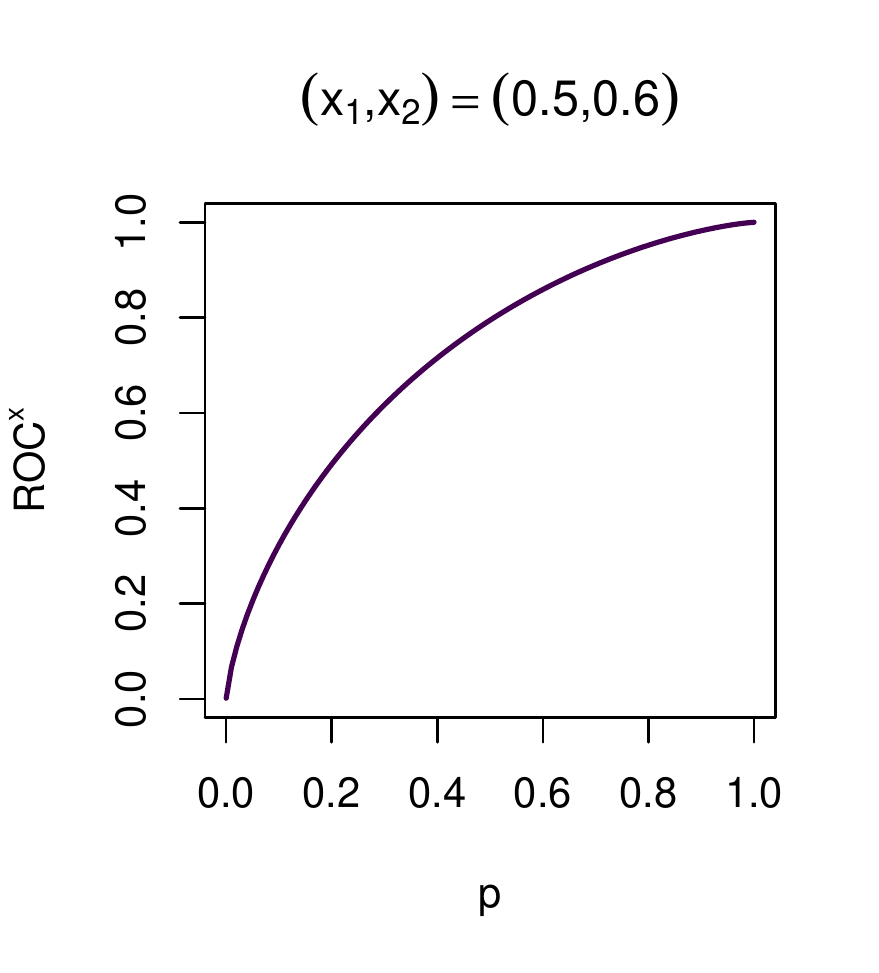}
&
\includegraphics[scale=0.65]{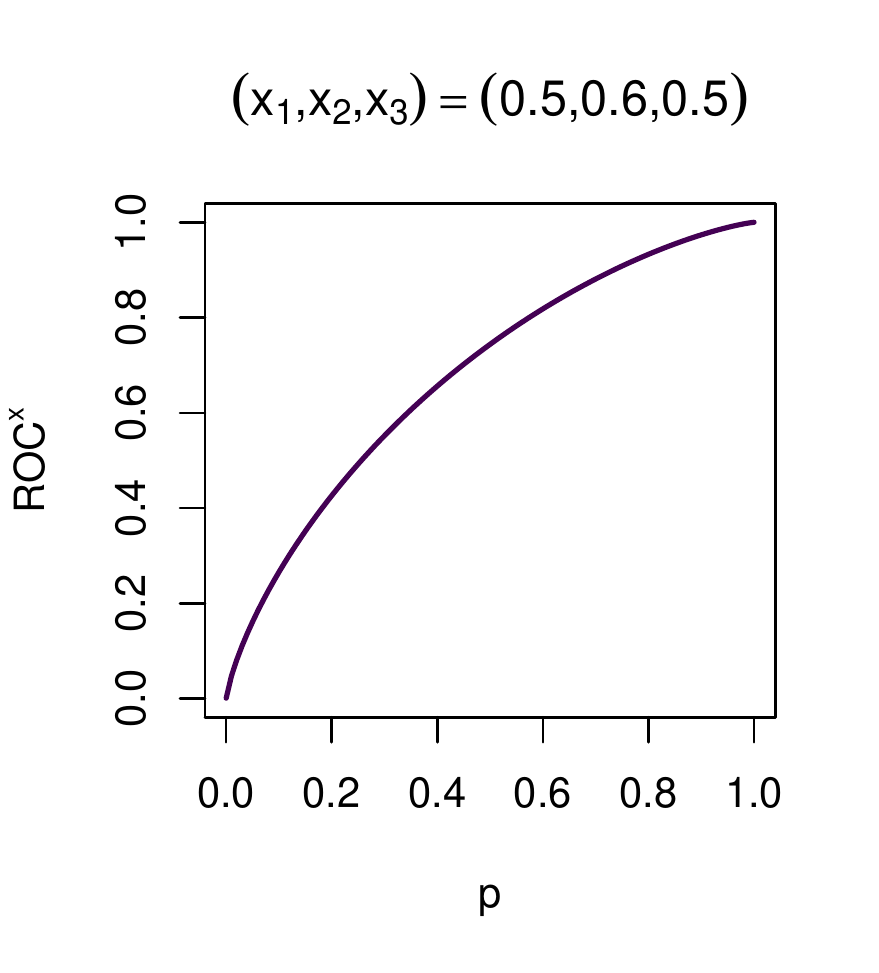}\\
\hline
\end{tabular}
\end{center}
\caption{Scenarios under the null hypothesis considered for calibrating the level of the test. }
\label{table:level}
\end{table}
The results of the simulations obtained for $n_{\bm{\beta}} = 5$ 
 are summarized in Figures \ref{fig:levelX2n5} (for $d = 2$) and  \ref{fig:levelX3n5} (for $d = 3$). 
 Each subfigure represents the test of one scenario for a particular sample size. The nominal level considered is 0.05. 
 The estimated proportion of rejections over 500 
replications of the data sets is represented along with the rejection region of such nominal level. For the test to be well calibrated the estimated proportions should fall between the gray lines.

In general it can be said that the expected nominal level is reached, as  most of the estimated proportions are close to the corresponding nominal level. The $L_2$ statistic seems to overestimate the level in a few scenarios, but its behaviour improves when increasing the sample size. The $KS$ statistic is a little more conservative.

\subsection{Power of the test}
\label{sec3.2}

On the other hand, the scenarios that were considered for studying the power of the test (by comparing different conditional ROC curves) are represented in Table~\ref{table:power}. 

\begin{table}[!t]
\begin{center}
\begin{tabular}{ m{1.2cm}  m{6cm} m{6.5cm}}
\hline
\centering   & \centering \textbf{$2-$dimensional covariate} &   \textbf{$3-$dimensional covariate}\\
\hline
\centering $K=2$ & \centering  $ROC_1^x \text{ vs. } ROC_2^x$ &  \qquad \quad $ROC_4^x \text{ vs. } ROC_5^x$ \\
&
\vspace{0.1cm}
\includegraphics[scale=0.65]{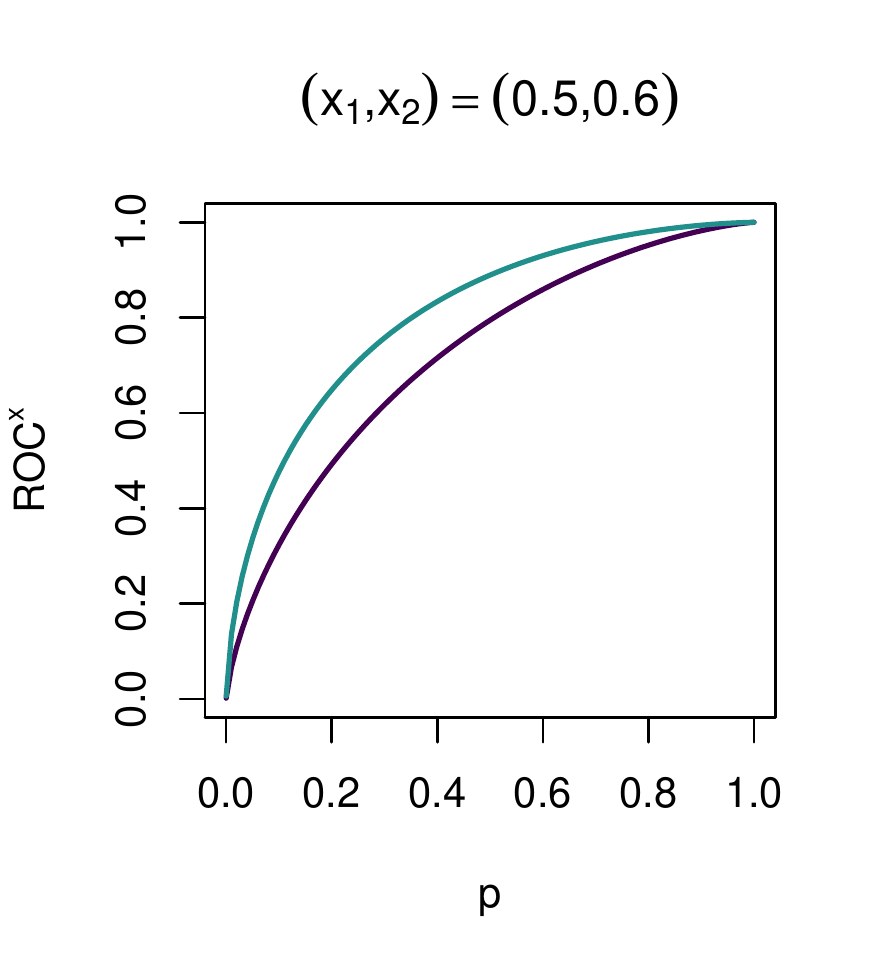}
&
\includegraphics[scale=0.65]{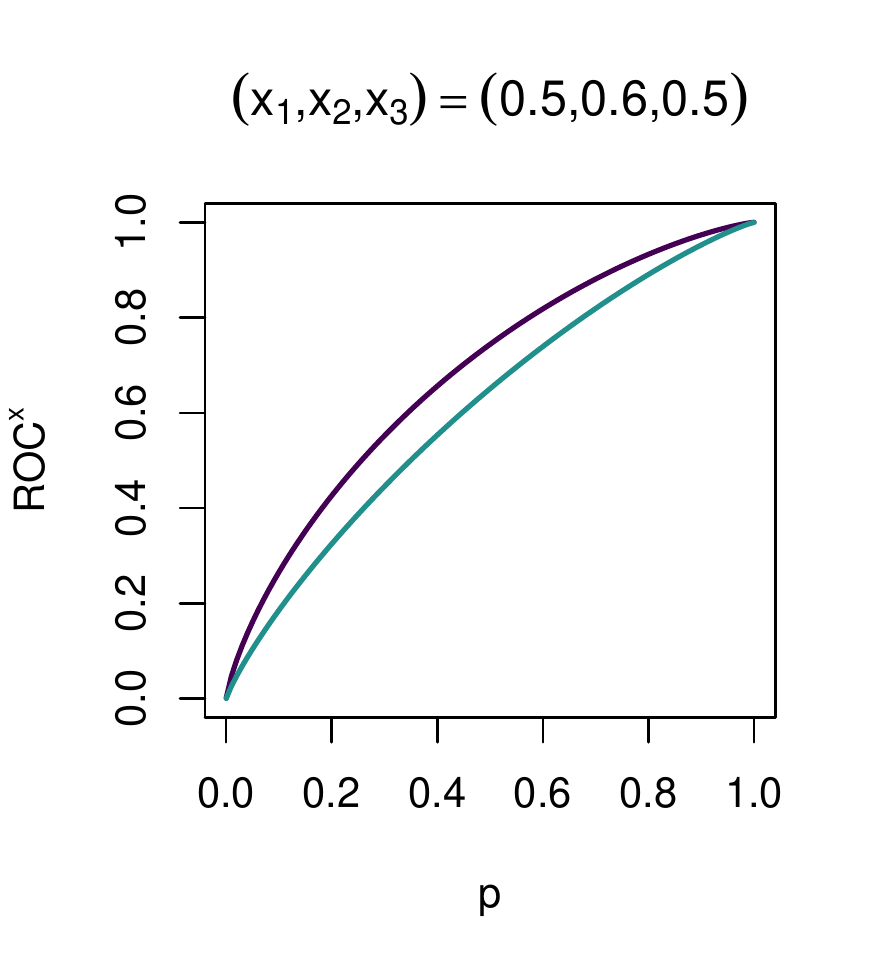}\\
\hline
\centering $K=3$ &  \centering $ROC_1^x \text{ vs. } ROC_2^x \text{ vs. } ROC_3^x$ &  \quad $ROC_4^x \text{ vs. } ROC_5^x \text{ vs. } ROC_6^x$\\
&
\vspace{0.1cm}
\includegraphics[scale=0.65]{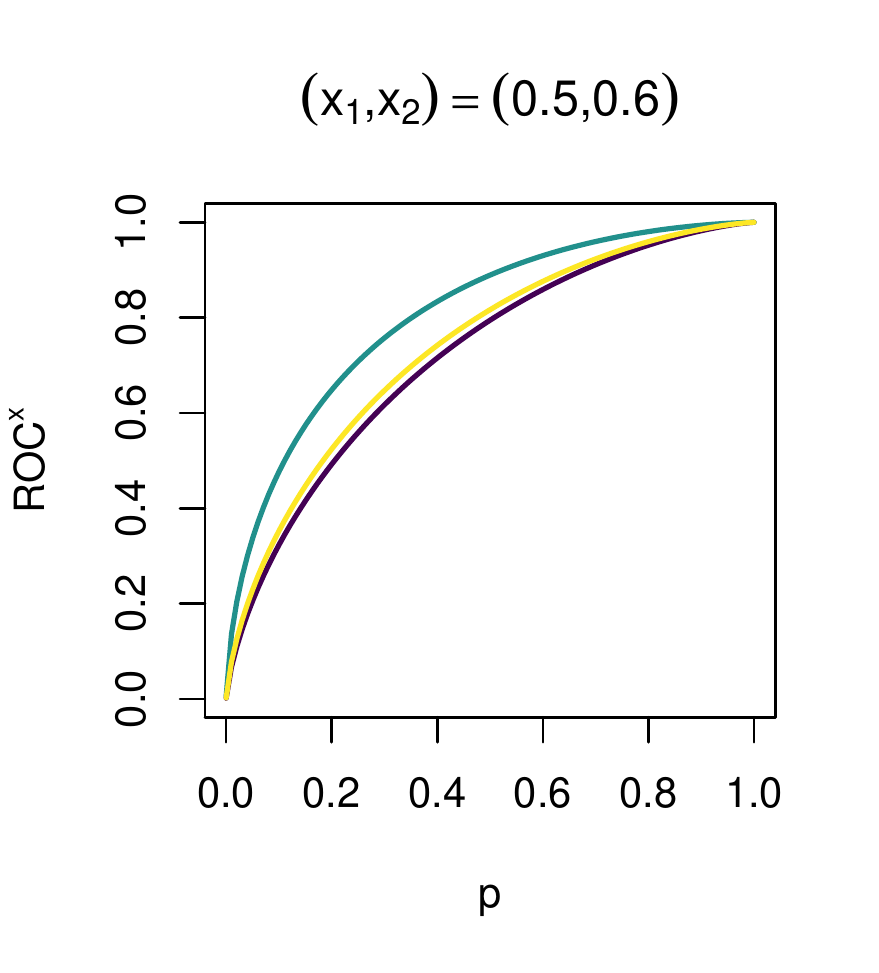}
&
\includegraphics[scale=0.65]{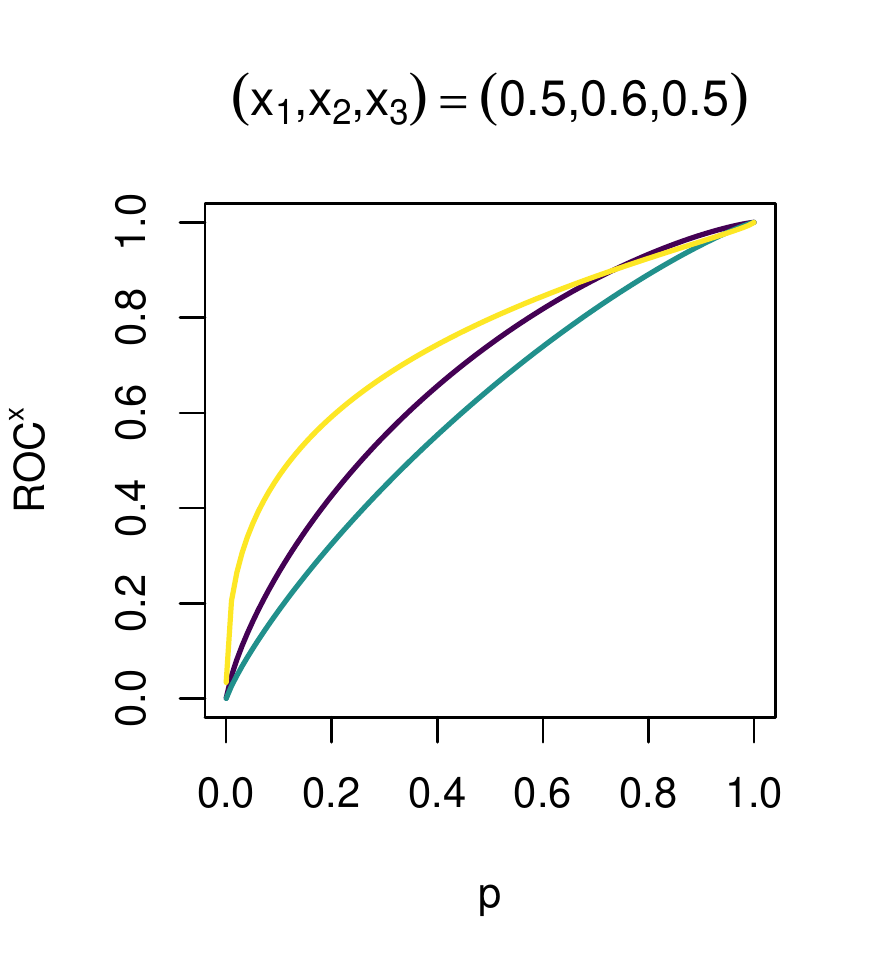}\\
\hline
\end{tabular}
\end{center}
\caption{Scenarios under the alternative hypothesis considered for calibrating the power of the test. $ROC_1^x$ and $ROC_4^x$ are represented in purple,  $ROC_2^x$ and $ROC_5^x$ in green, and $ROC_3^x$ and $ROC_6^x$ in yellow.}
\label{table:power}
\end{table}

\begin{figure}[!b]
\centering\includegraphics[scale=0.49]{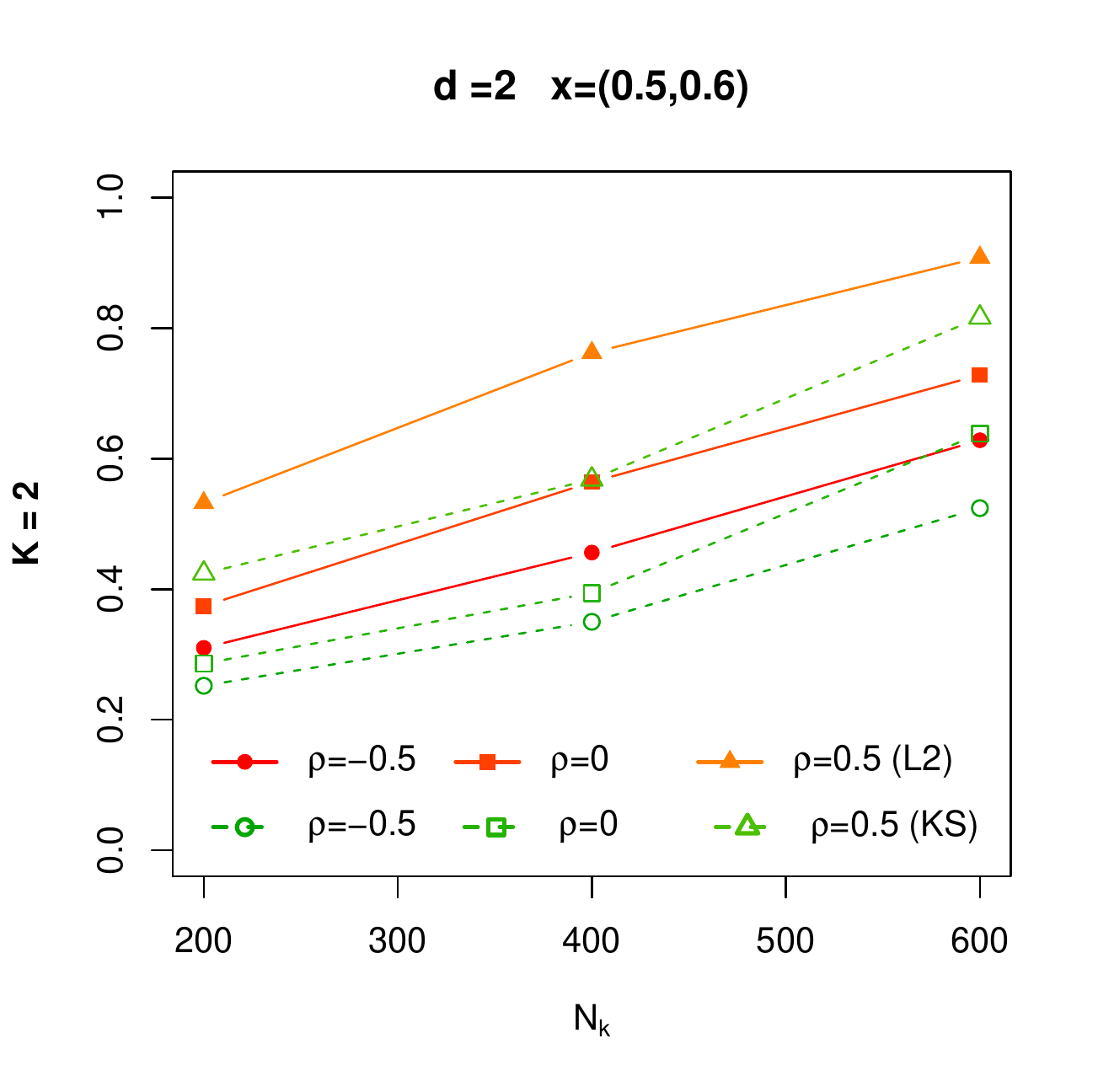} 
\centering\includegraphics[scale=0.49]{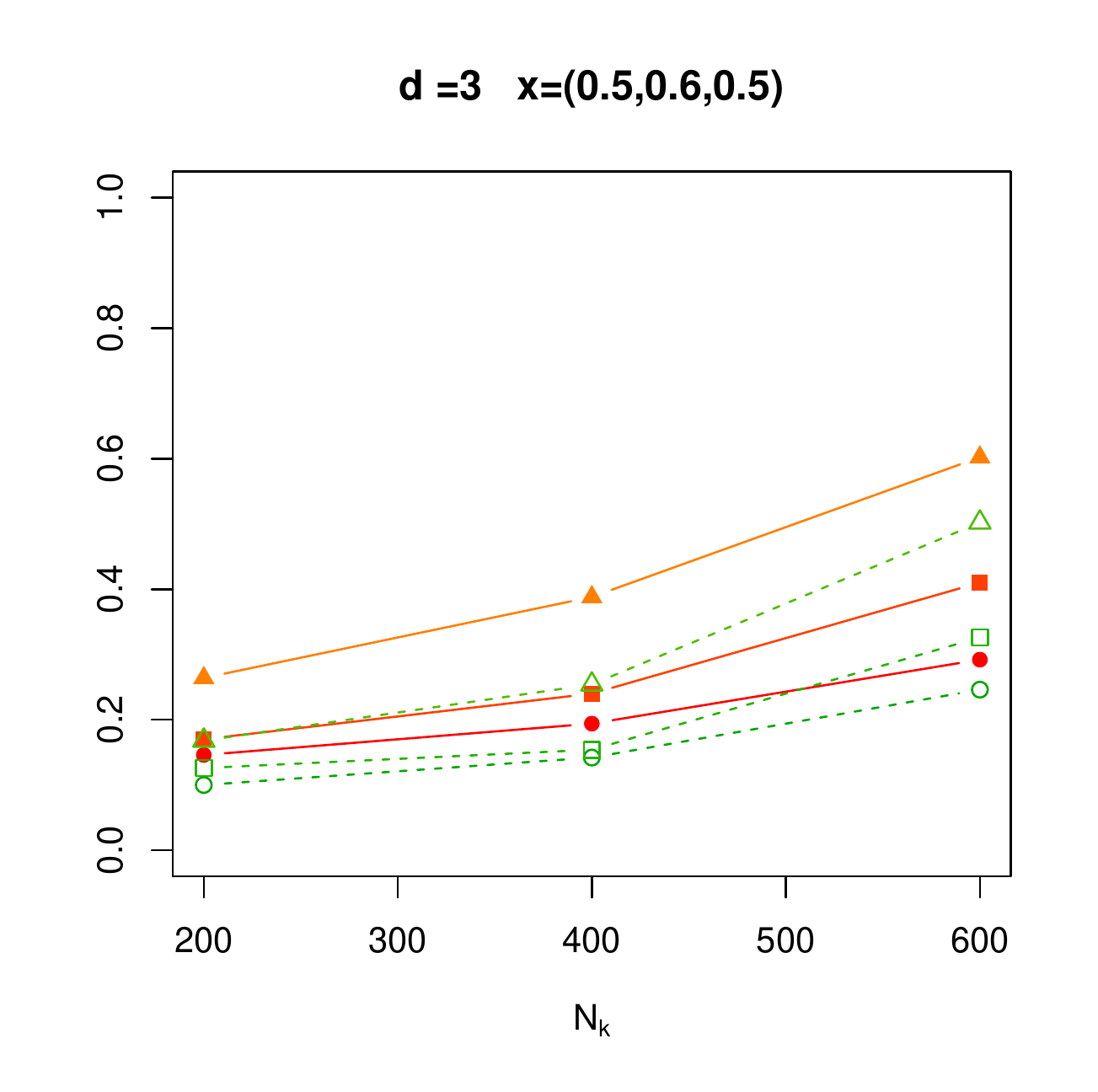}

\centering\includegraphics[scale=0.49]{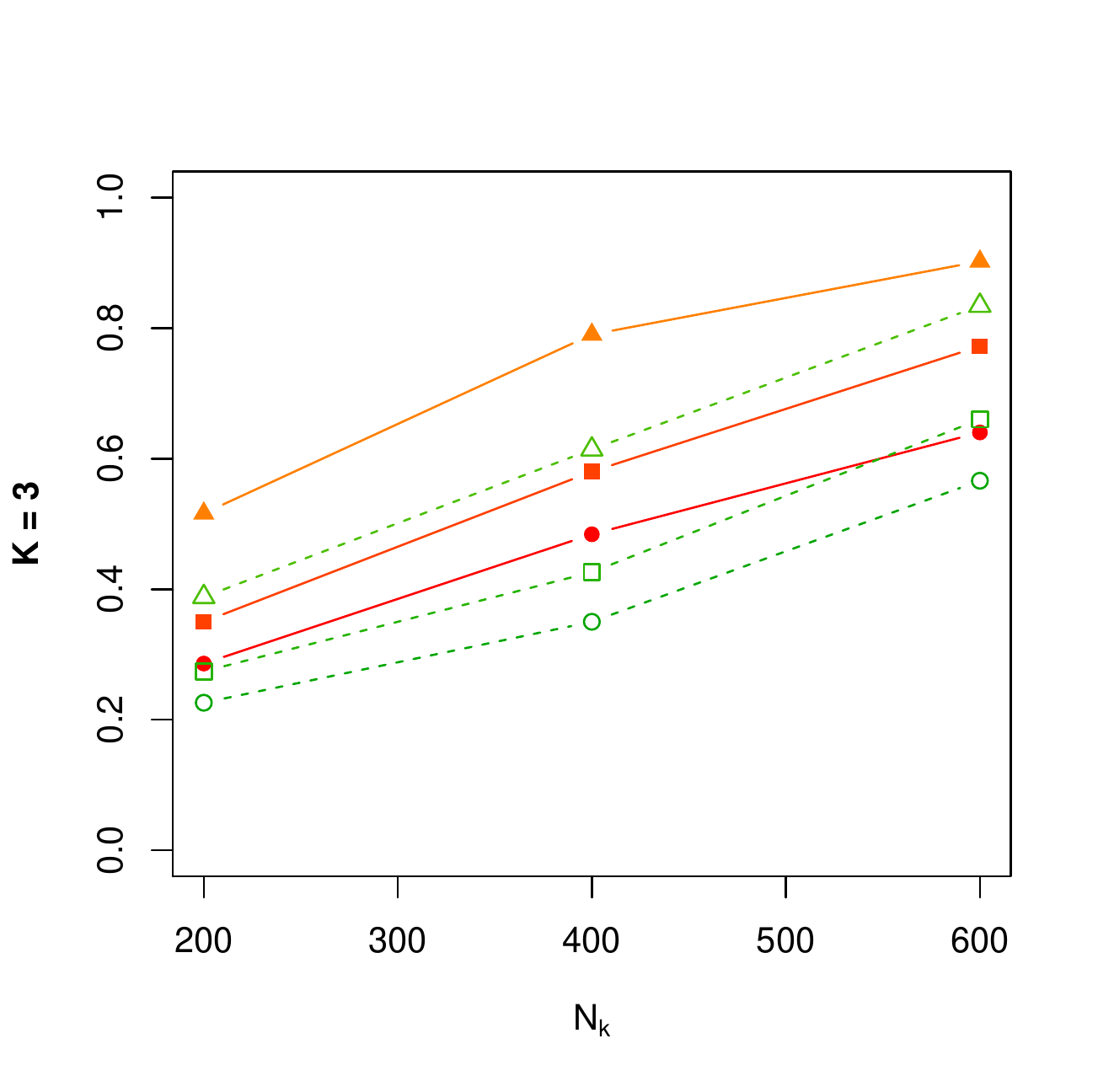}
\centering\includegraphics[scale=0.49]{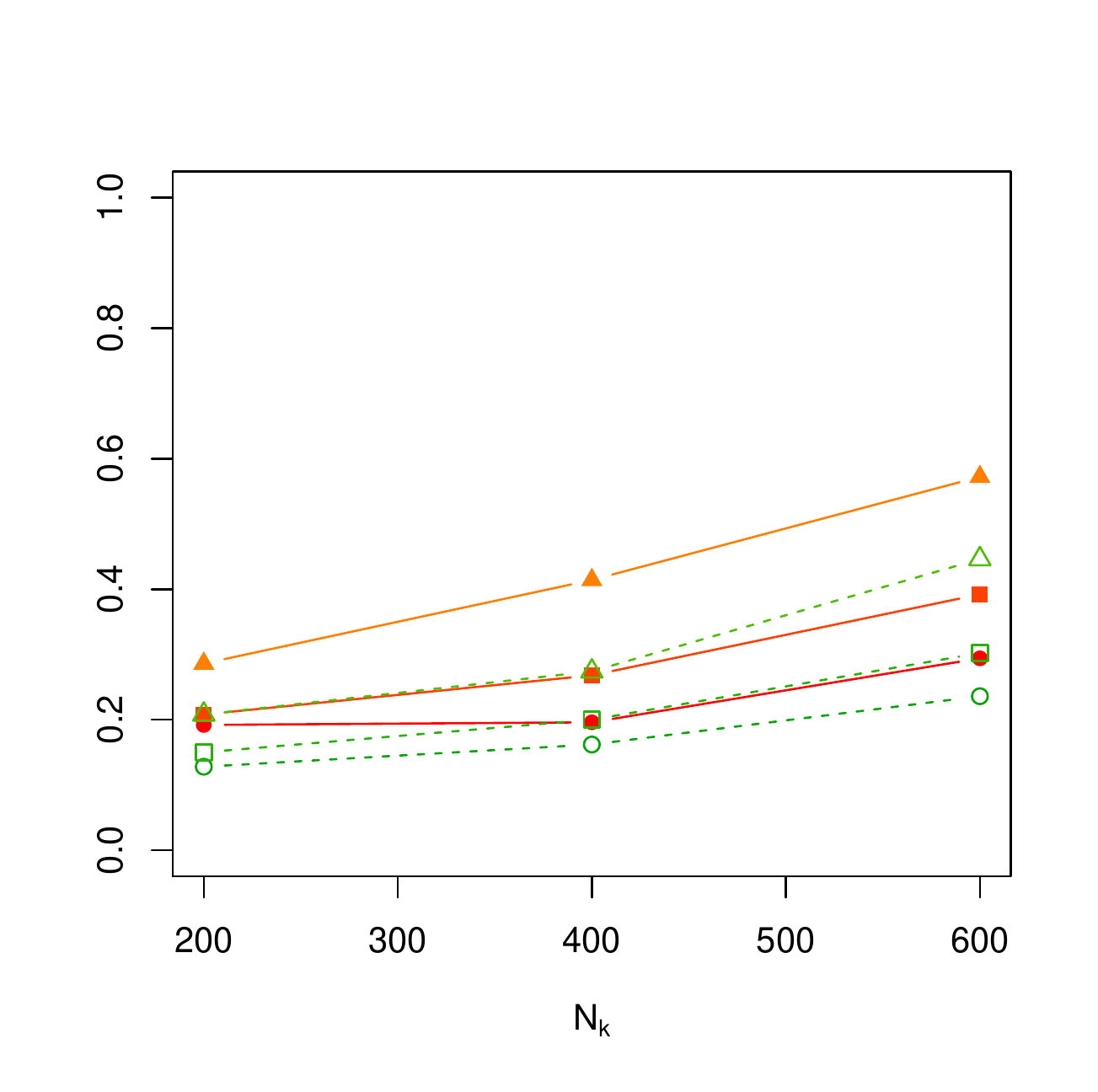}
\caption{Estimated proportion of rejection under the alternative hypothesis for different sample sizes and different $\rho$, for $n_{\bm{\beta}}=5$ ($\alpha = 0.05$).}
\label{fig:powern5}
\end{figure}

The results of the simulations are summarized in Figures \ref{fig:powern5} (for $n_{\bm{\beta}}=5$). In those figures the first and second row represent the simulation results for the scenarios with $K=2$ and $K=3$, respectively, and the first and the second column represent the simulation results for $d=2$ and for $d=3$, respectively. In this case,  only $\alpha = 0.05$ was considered.

It can be seen that the power of the test grows with the considered sample sizes. The $L_2$ statistic yields  higher power than the $KS$ statistic, which is consistent with $KS$ being more conservative. Moreover, the difference between the conditional ROC curves considered for the case of $d=2$ is bigger than the difference between the ROC curves in the scenarios with $d=3$, which translates in higher power for the  cases in which $d=2$.

We can also observe that for each scenario, the highest power is always obtained for the cases in which the correlation of the diagnostic variables is $\rho = 0.5$, and the lowest for $\rho = -0.5$. 

{ Note that for the scenario with $d=3$ and $\rho=-0.5$ the power of the test does not increase significantly from the first sample size to the second (in fact, for $K=3$ it even decreases a little), but this can be due to the fact that the lower sample size has balanced data, ($n^F$,$n^G$) being (100,100). whereas for the second sample size considered ($n^F$,$n^G$) take the value (250,150). The highest sample size is also unbalanced, but not so much.}

\begin{remark}
In order to evaluate the modification of the method proposed in Remark \ref{remark1} and \ref{remark2} we have run simulations for the same scenarios previously described. We show here the results for the scenarios with $K=2$ and $d=2$ under the null and the alternative hypotheses  for assessing  the level and the power of the test, respectively. Similar conclusions were obtained with  the rest of the scenarios. {The  parameters that are used here are the same as before, with the exception that now 1000 data-sets were simulated instead of 500.}

Figure \ref{fig:levelremark} shows the results of the simulations when considering the modification of Remark \ref{remark1} for $m_{\bm{\beta}} =50$ (first row) and $m_{\bm{\beta}} =25$ (second row), and the results for considering only one random projection (Remark~\ref{remark2}), i.e., $m_{\bm{\beta}} =n_{\bm{\beta}}=1$ (third row). Note that taking $m_{\bm{\beta}} =25$ is comparable with $n_{\bm{\beta}} =5$ used in the previous simulations (see first row of Figure~\ref{fig:levelX2n5}), and that the results are very similar: the estimated proportion or rejections is a little overestimated for the $L_2$ statistic for the smaller sample size and otherwise close to the nominal level, and the $KS$ statistic is always more conservative. 
Increasing $m_{\bm{\beta}}$ from 25 to 50 does not seem to affect the results significantly, and neither does reducing it to a single random projection ($m_{\bm{\beta}} =1$).
\begin{figure}[!h]
\centering\includegraphics[scale=0.41]{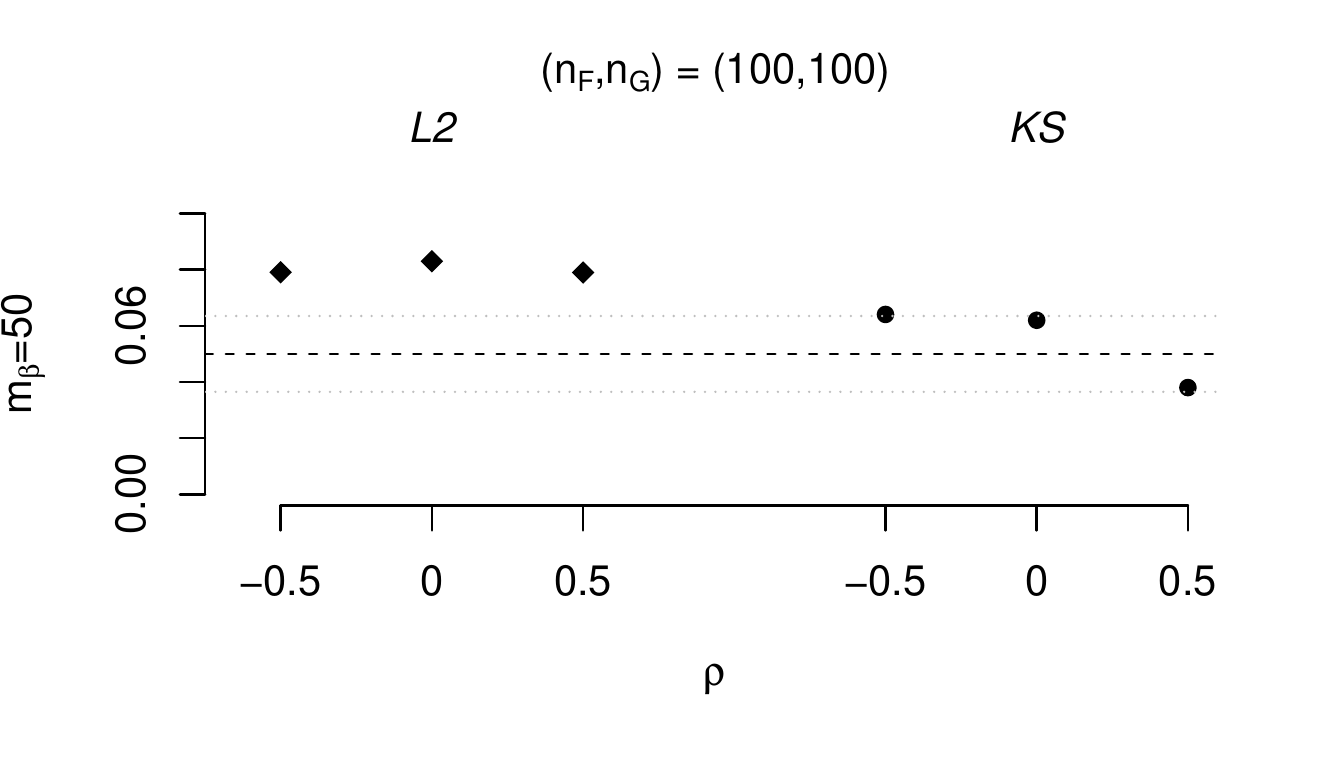}
\centering\includegraphics[scale=0.41]{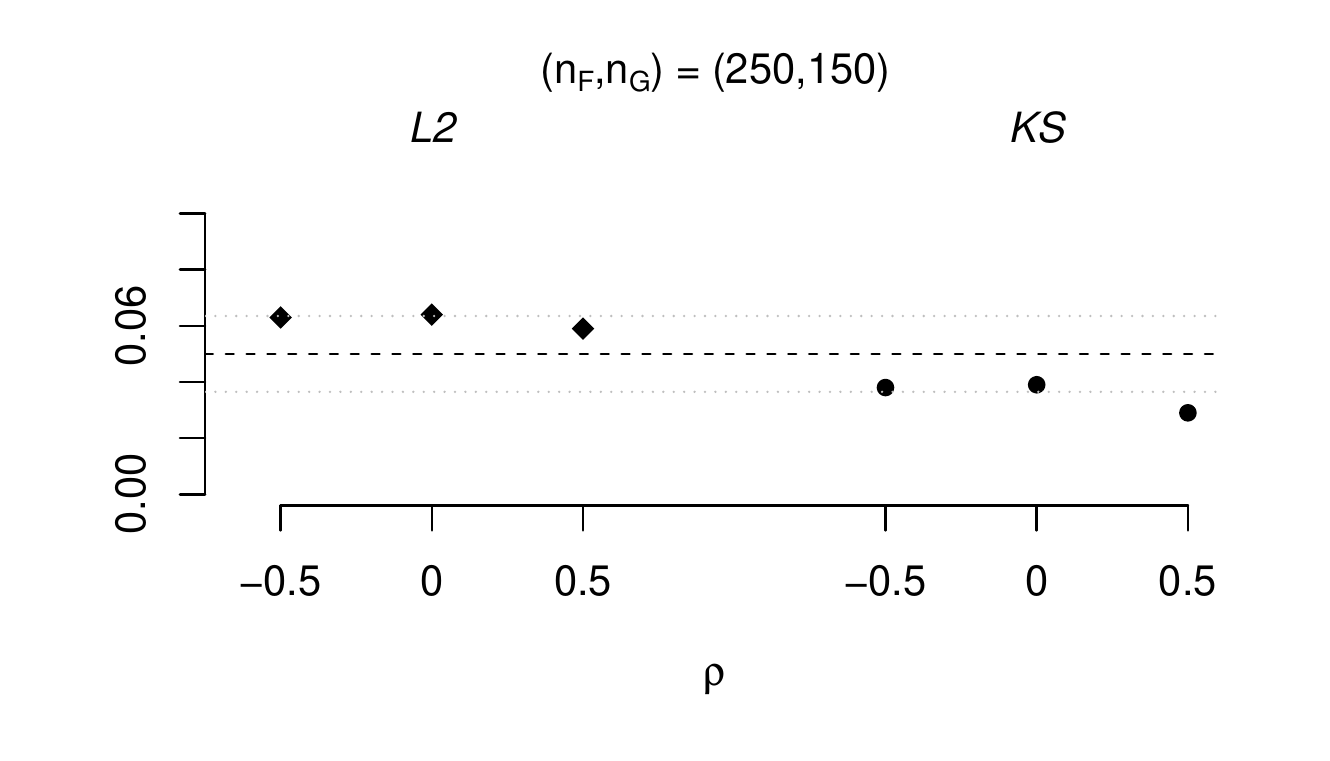}
\centering\includegraphics[scale=0.41]{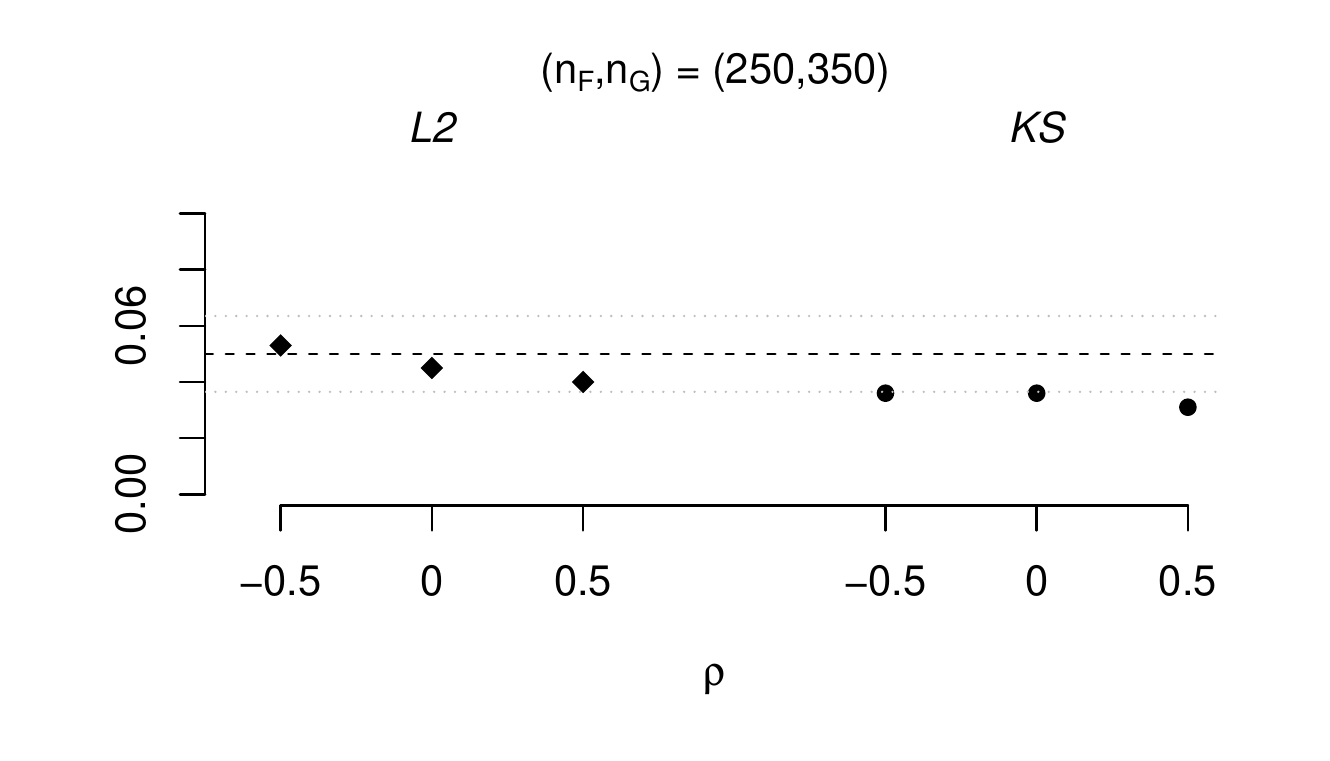}

\includegraphics[scale=0.41]{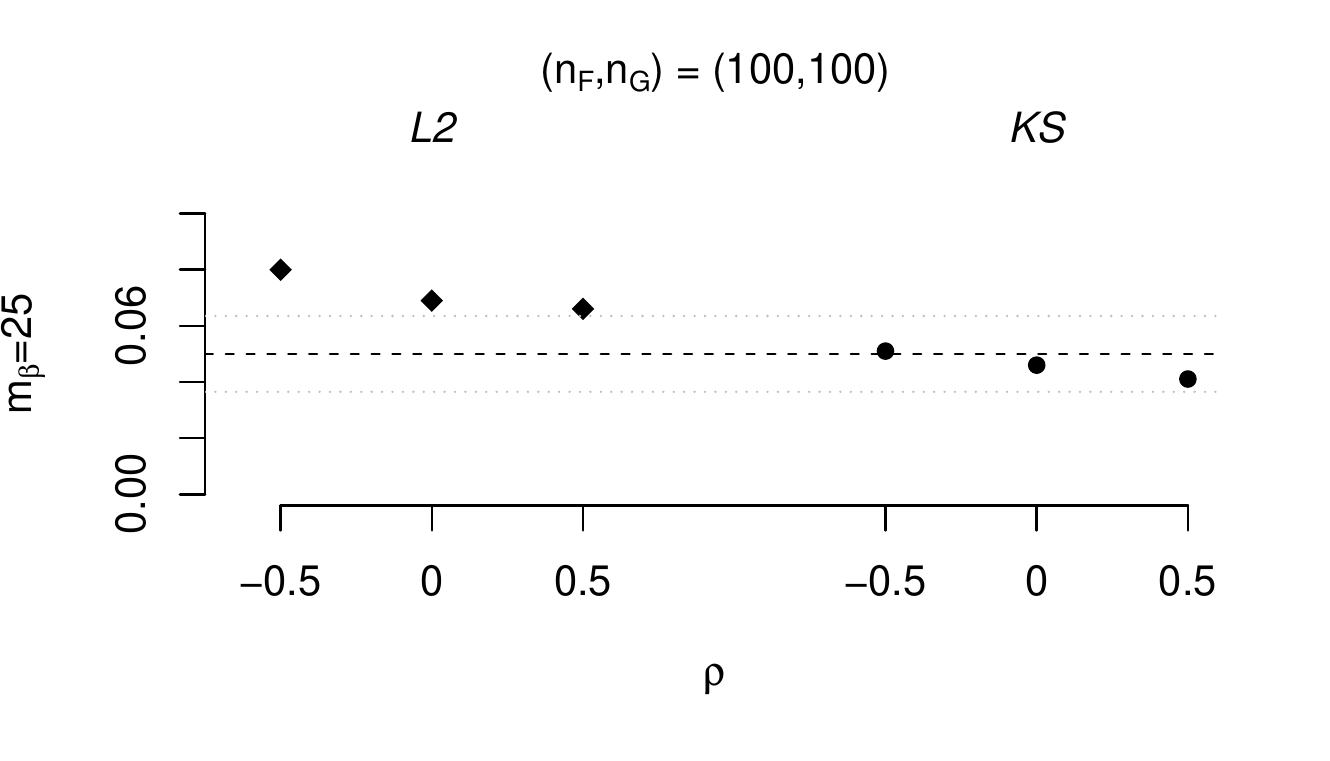}
\centering\includegraphics[scale=0.41]{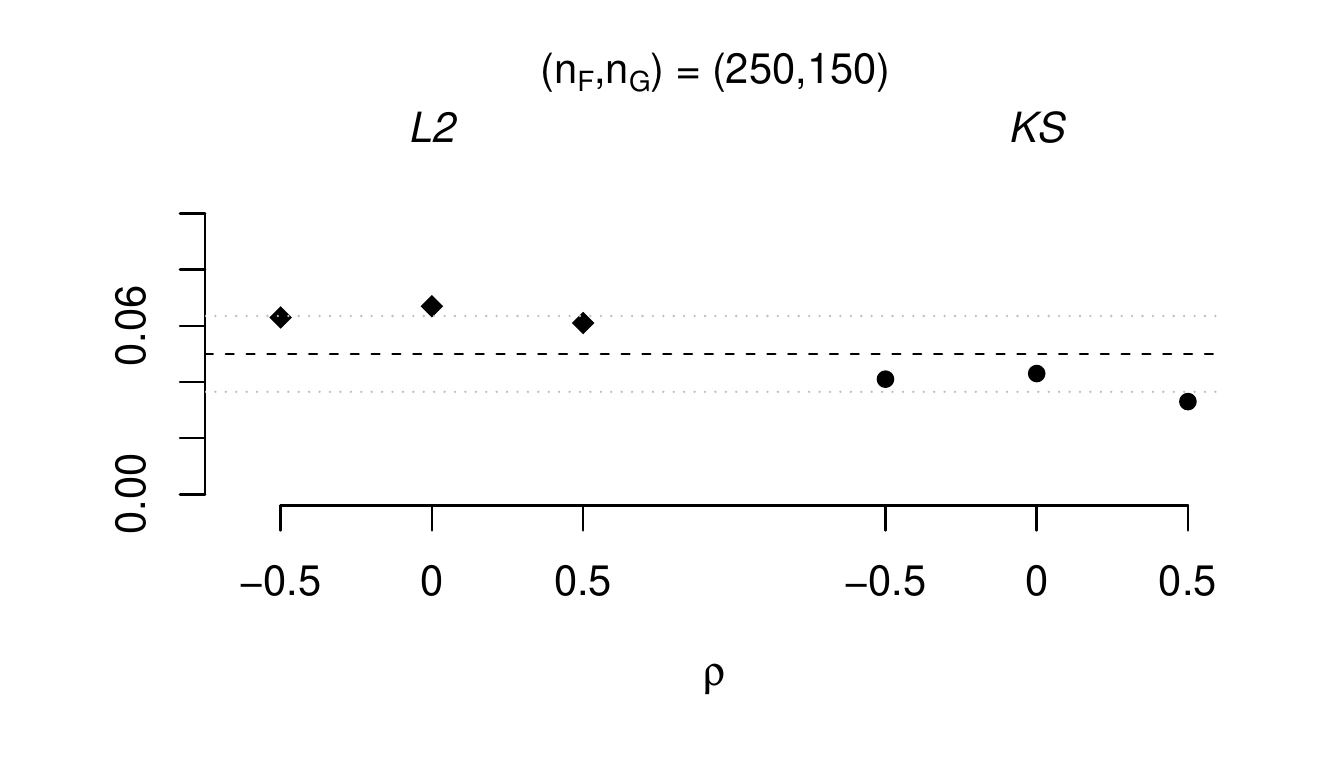}
\centering\includegraphics[scale=0.41]{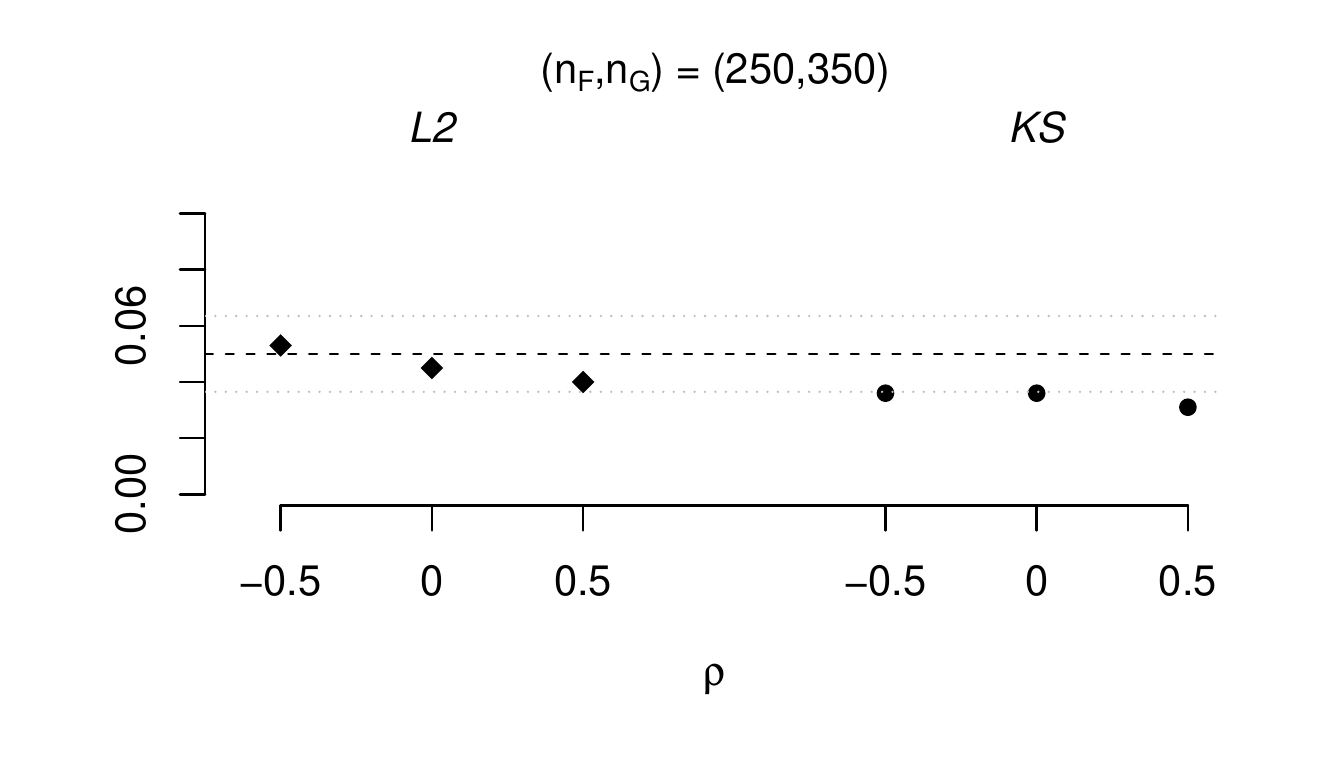}

\includegraphics[scale=0.41]{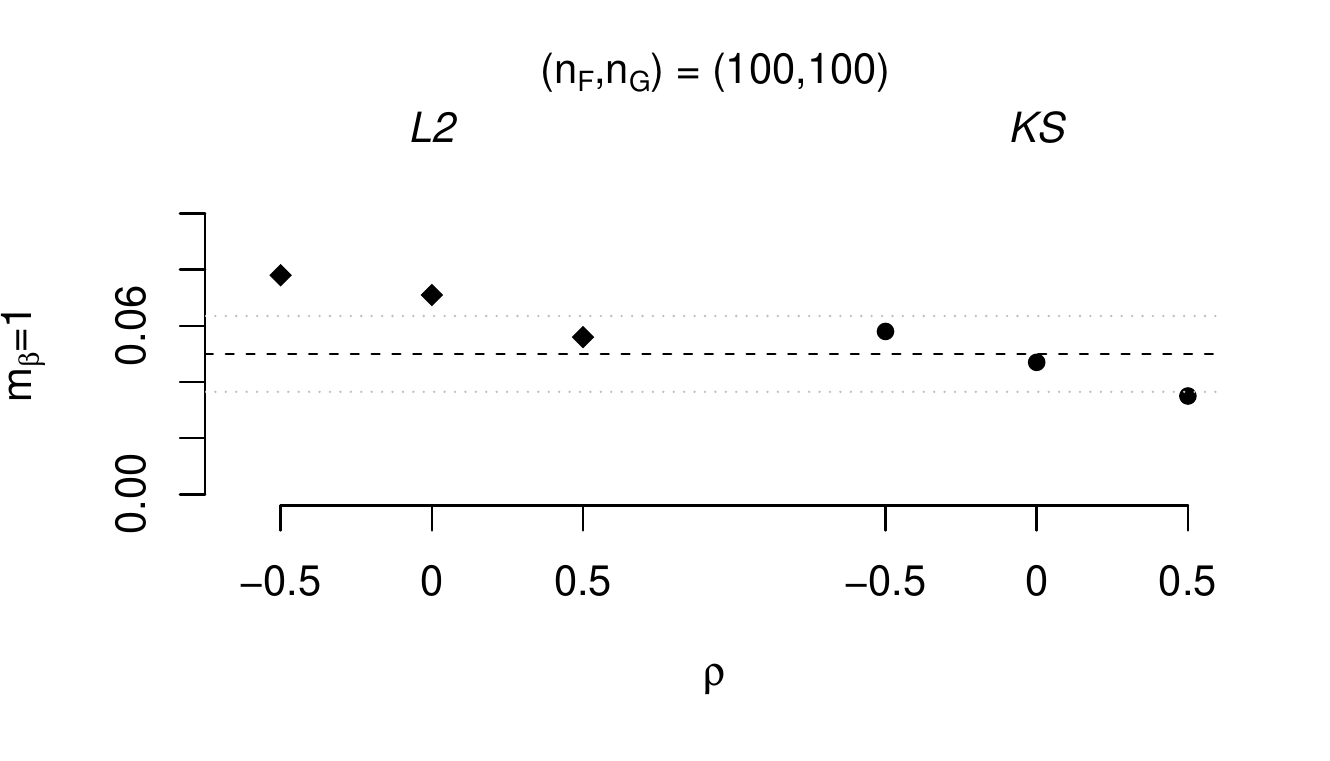}
\centering\includegraphics[scale=0.41]{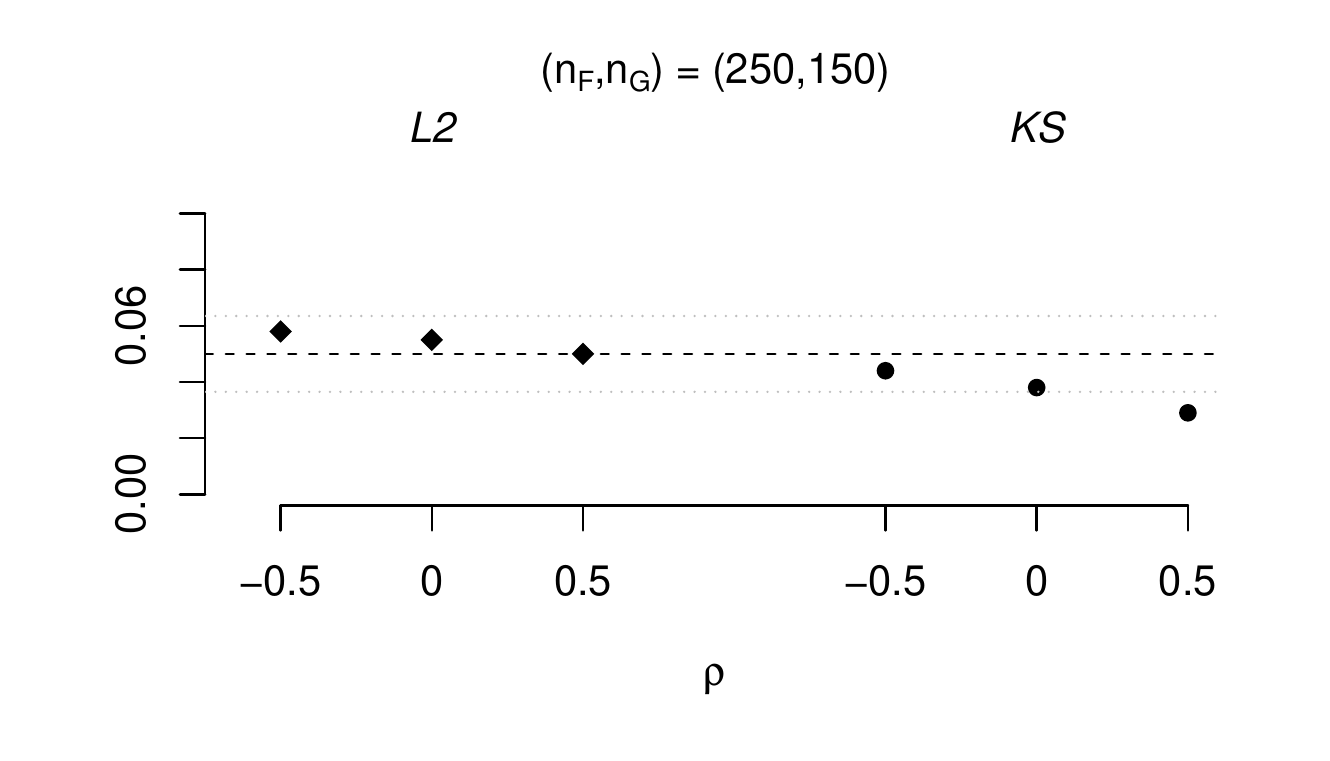}
\centering\includegraphics[scale=0.41]{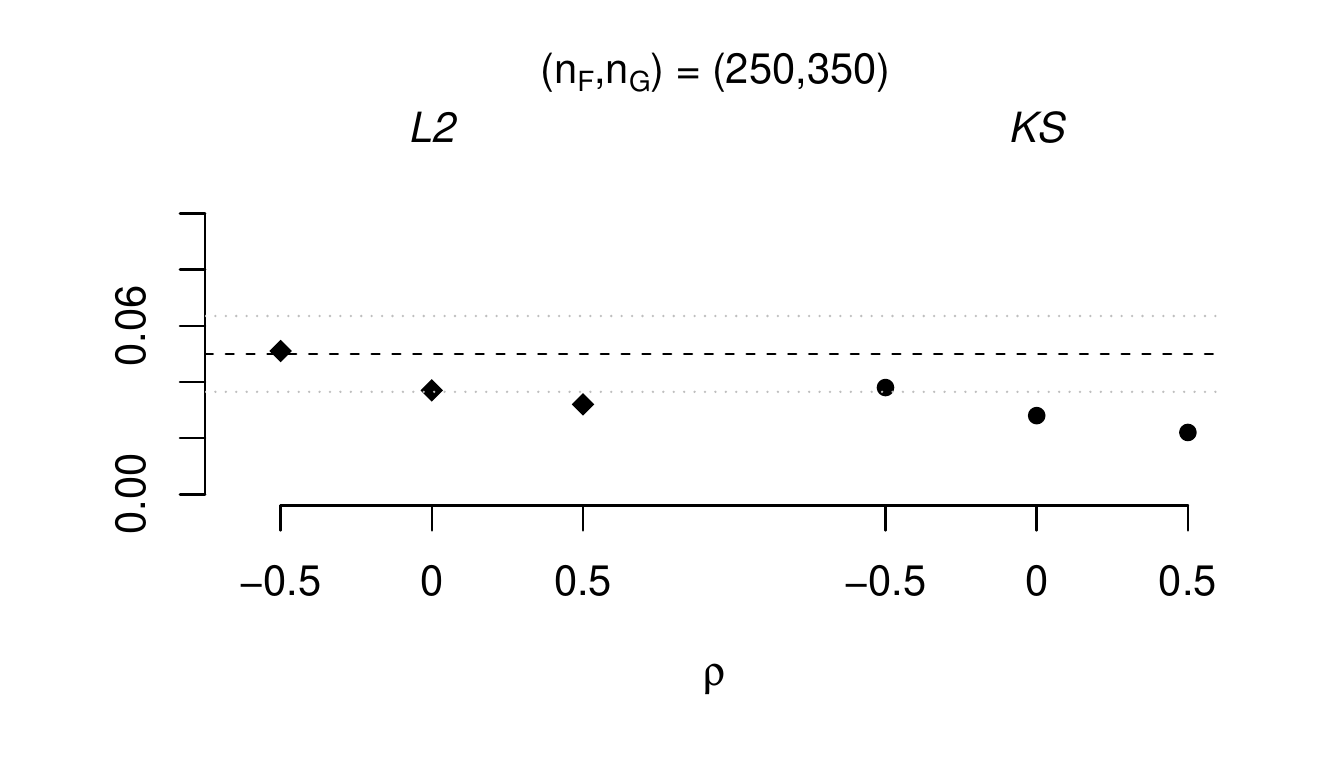}
\caption{Estimated proportion of rejection under the null hypothesis and the corresponding limits of the critical region (in gray) for the level 0.05 (dotted black line) with $K=2$, $d=2$ and $m_{\bm{\beta}} = 50, 25, 1$ for different sample sizes and different $\rho$.}
\label{fig:levelremark}
\end{figure}

In Figure \ref{fig:powerremark} we can observe the results for the simulations under the alternative hypothesis, once again for $m_{\bm{\beta}} =50$, $m_{\bm{\beta}} =25$ and $m_{\bm{\beta}} =1$. The firs two graphics are very similar to the one obtained for $n_{\bm{\beta}} =5$ (see the first graphic of Figure~\ref{fig:powern5}), but from the last graphic it is obvious that by using only one random projection the power of the test decreases considerably (as it was expected).

\begin{figure}[!]
\centering\includegraphics[scale=0.43]{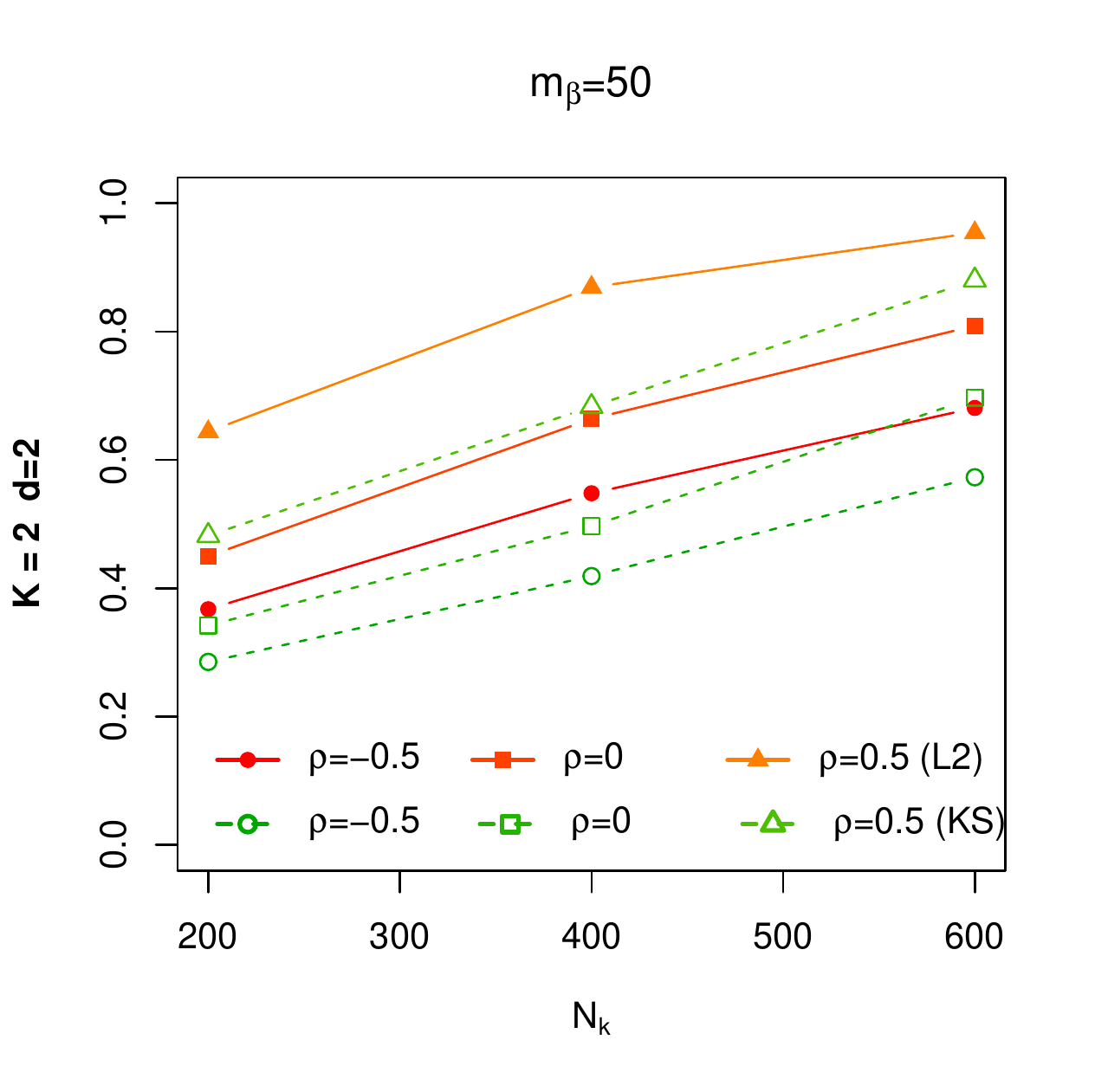} 
\centering\includegraphics[scale=0.43]{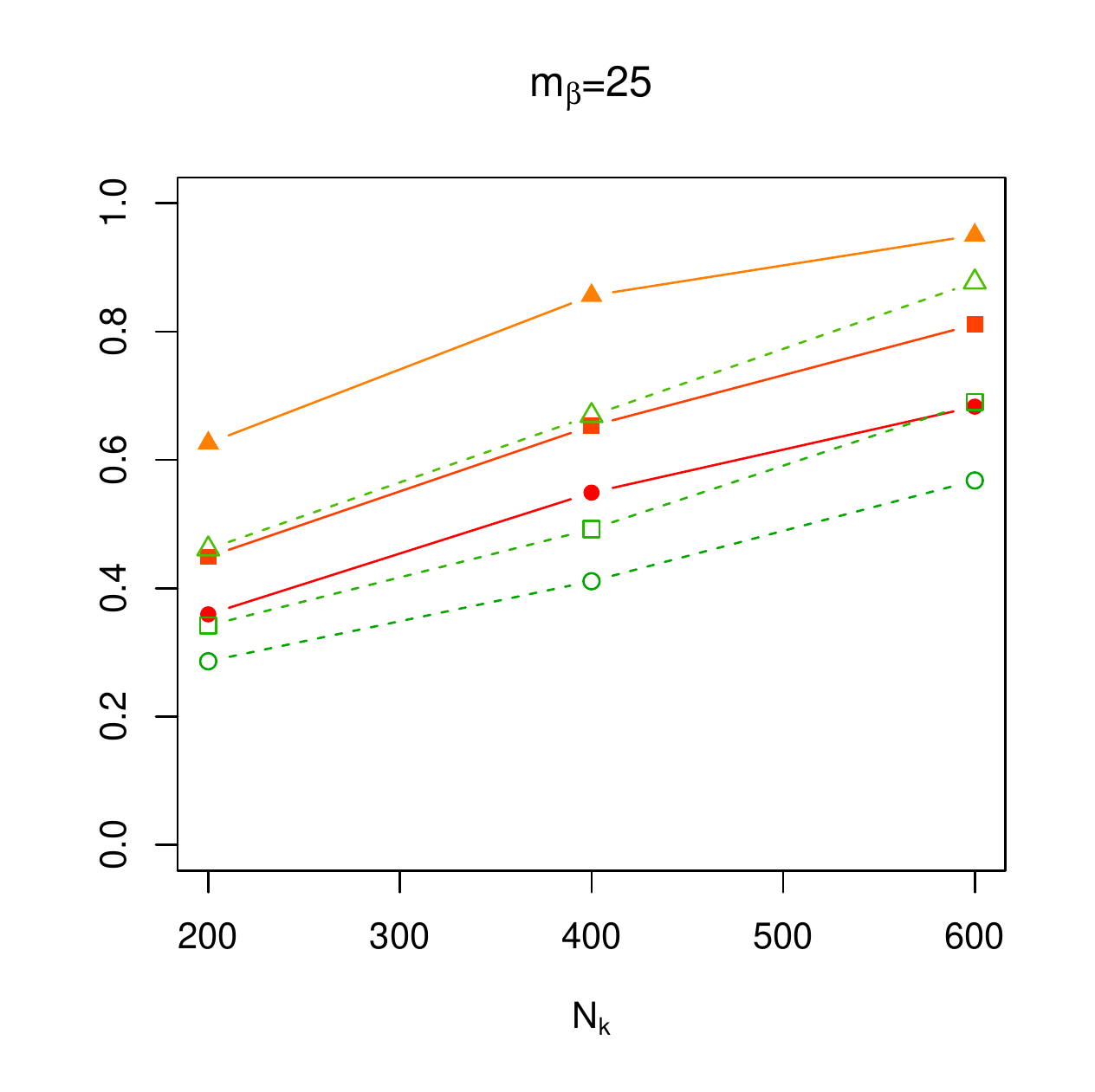}
\centering\includegraphics[scale=0.43]{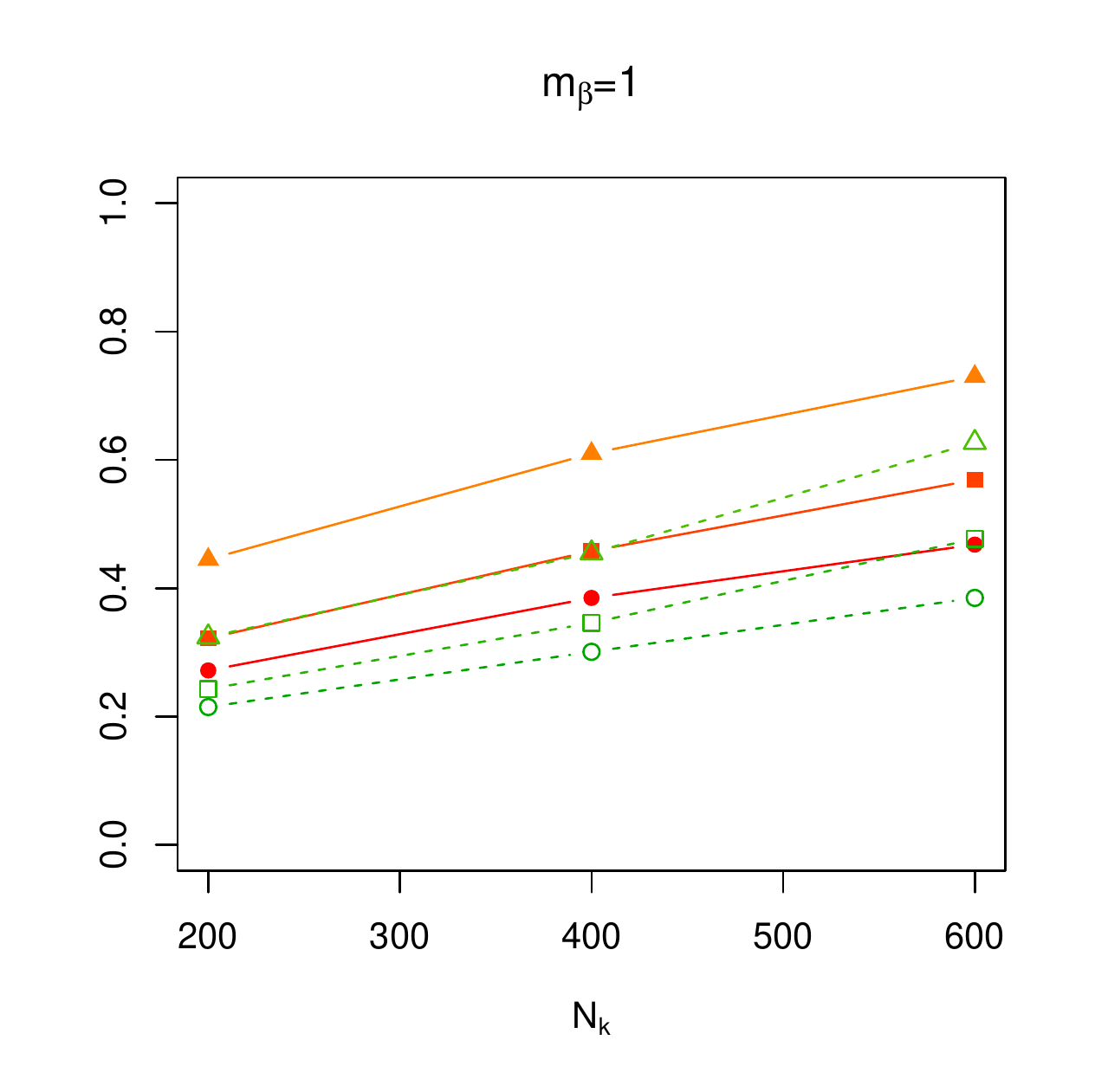}

\caption{ Estimated proportion of rejection under the alternative hypothesis for different sample sizes and different $\rho$, for $n_{\bm{\beta}}=50,25,1$ and for the scenarios with $K=2$ and $d=2$.}
\label{fig:powerremark}
\end{figure}

In the light of these results it seems that the alternative methodology proposed in Remark~\ref{remark1} yields similar conclusions than the first proposal, with no noticeable gain when increasing the number $m_{\bm{\beta}}$ used to approximate the value of the statistic from 25 to 50. It remains an open problem to determine an optimal value for that parameter. 

As for the idea mentioned in Remark~\ref{remark2}, using only one random projection seems to produce a well calibrated test, despite having considerably lower power.
%

\end{remark}

\section{Application}
\label{sec4}
An illustration of the proposed test is displayed in this section through the analysis of {\color{dgreen}the previously mentioned} data set concerning 463 patients with  pleural effusion. This data set has been provided by Dr. F. Gude, from the Unidade de Epidemiolox\'ia Cl\'inica of the Hospital Cl\'inico Universitario de Santiago (CHUS), and it has been used for a previous study in \cite{Valdes2013}.

\begin{figure}[!ht]
\centering\includegraphics[scale=0.6]{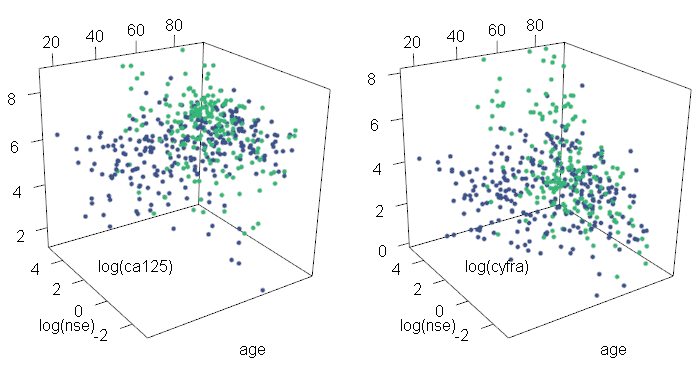} 
\caption{Scatterplot of the three different diagnostic biomarkers in function of the two covariates considered: \textit{age} and ${\log(nse)}$. The healthy subjects are represented in blue and the diseased ones in green.}
\label{fig:appli}
\end{figure}

From a medical perspective, the goal is to find a way to discriminate the patients in which the pleural effusion (PE) has a malignant origin (MPE) from those in which the PE is due to other non-cancer-related causes. 200 individuals form the sample had MPE (the diseased population in this context), against 263 who did not (healthy population). 
For that matter, two diagnostic markers were considered, the carbohydrate antigen 152 (\textit{ca125}) and the cytokeratin fragment 21-1 (\textit{cyfra}). Moreover, the information of two different covariates is also available: the \textit{age} and the neuron-specific enolase (\textit{nse}). Due to the characteristics of the data (positive values, most of them close to zero, with some extreme high values), logarithms of those variables -- excluding the variable \textit{age} -- were considered for the study. Being the logarithm a monotone transformation, its use does not have an effect on the estimation of the common ROC curve. However, it does affect the estimation of the conditional ROC curves, as it reduces the effect of the more extreme values of the variables. A representation of the relationship of each one of those biomarkers with the two covariates is depicted in Figure~\ref{fig:appli}, for both MPE (green) and the non-MPE (blue) patients.
It can be observed that the shape of the point clouds of the two populations  changes with the values of the covariates, specially in the case of the diseased population.

In order to evaluate whether the discriminatory capability of those markers 
($Y_1^F$ and $Y_1^G$ as the variables containing the information of $\log(ca125)$, and $Y_2^F$ and $Y_2^G$ as the variables containing the information of $\log(cyfra)$) 
is the same when the covariates \textit{age} and ${\log(nse)}$ are taken into account, the methodology explained in previous sections is applied, comparing their respective ROC curves conditioned to different values of the bidimensional covariate $\bm{X} = (X_1, X_2)$ with $X_1 =$ \textit{age} and $X_2 ={\log(nse)}$. In order to explore the advantages of using this method over the ones that do not consider multidimensional covariates, we also test the equivalence of the ROC curves of those diagnostic markers for the case in which no covariates are taken into account and for the case in which only one of the covariates is included in the analysis. 

\begin{figure}[!ht]
\centering\includegraphics[scale=0.5]{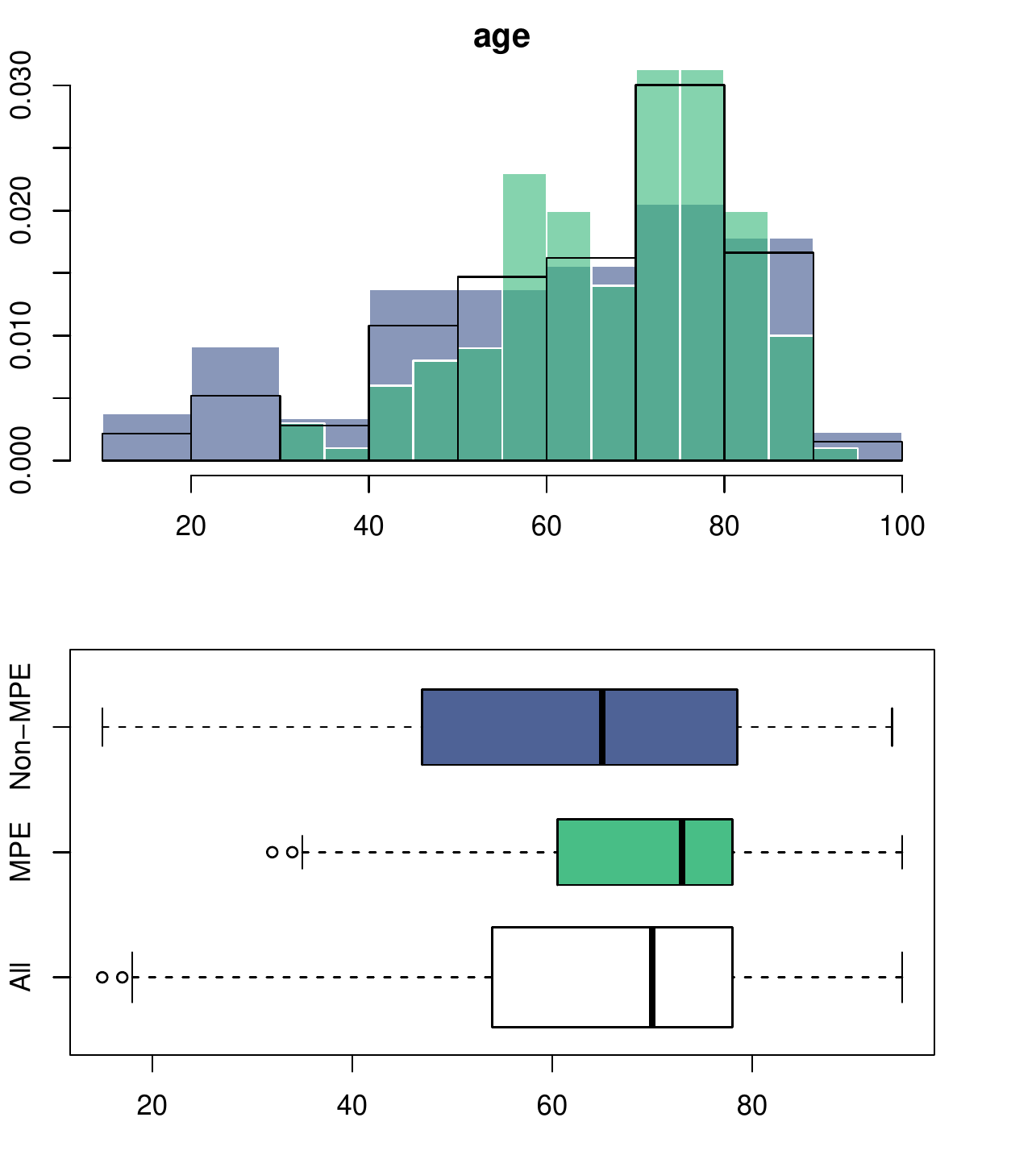} 
\centering\includegraphics[scale=0.5]{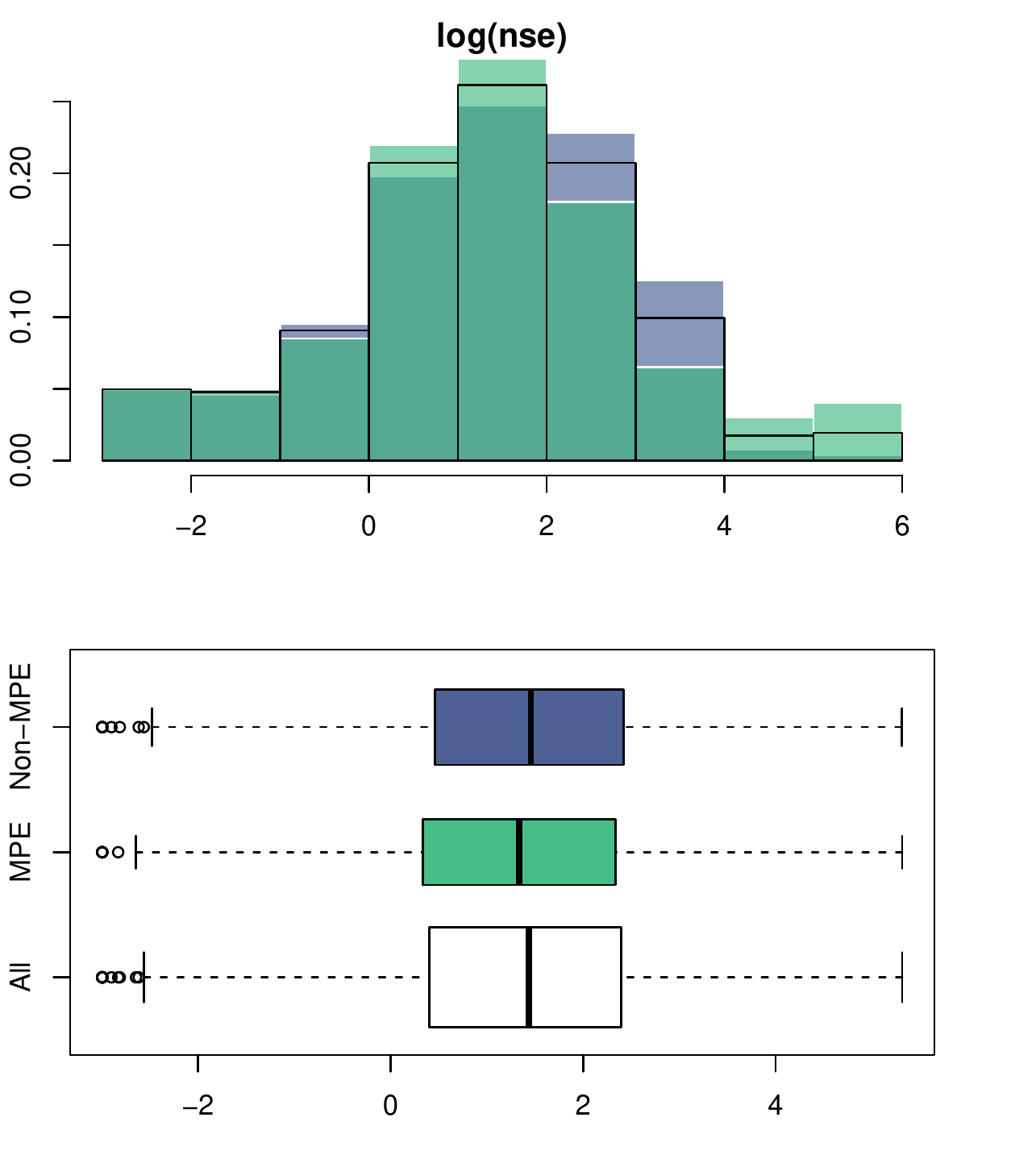} 
\caption{Histograms and boxplots of the two covariates considered (\textit{age} and $\log(nse)$). The healthy subjects are represented in blue and the diseased ones in green. The black histogram lines and the white boxplot correspond to the two populations of the healthy and the diseased patients combined. }
\label{fig:histandbox}
\end{figure}

Figure~\ref{fig:histandbox} shows how those two covariates are distributed in the diseased and healthy populations.
Note that the covariates have different magnitudes: the values that the variable \textit{age} takes are always going to be bigger than the values of ${\log(nse)}$. Thus, if we were to use the procedure directly over these variables, when projecting the multidimensional covariate $\bm{X}$ on any direction, the effect of the second component will be overshadowed by the first component's. To prevent this from happening we decided to use the standardized variables of $X_1$  and $X_2$ instead of the originals. This also affects the value $\bm{x}$ at which the conditional ROC curves are being compared. { Note that an ROC curve conditioned to a certain value $x$ is the same as the  ROC curve in which the covariate is modified by a one-to-one transformation and that is conditioned to the corresponding  transformed $x$ value.}
 
 {
 Given a non-degenerate multidimensional covariate $\bm{X}$ the standardization proposed here is to consider the multidimensional covariate $\bm{X}_s = \bm{B}^{-1}(\bm{X} - \bm{a})$, with $\bm{B}$  a diagonal matrix with \\ $(\sqrt{Var(X_1)}, \dots, \sqrt{Var(X_d)})$ in the diagonal and $\bm{a}=(E(X_1),\cdots, E(X_d))'$. Then, for a given variable $Y$, a given $y\in \mathbb{R}$ and a certain value of the covariate $\bm{x}$, 
\begin{eqnarray*}
P(Y\leq y | \bm{X} = \bm{x})= P(Y\leq y |\bm{B}^{-1}(\bm{X} - \bm{a}) = \bm{B}^{-1}(\bm{x} - \bm{a})) = P(Y\leq y |\bm{X}_s = \bm{x}_s)),
\end{eqnarray*}
with $\bm{x}_s=\bm{B}^{-1}(\bm{x} - \bm{a})$ and, thus,
\begin{eqnarray*}
ROC^{\bm{x}}(p) = 1-F(G^{-1}(1-p| \bm{x})|\bm{x}) = 1-F(G^{-1}(1-p| \bm{x}_s)|\bm{x}_s) = ROC^{\bm{x}_s}(p)  ,
\end{eqnarray*}
Note that the standardization that takes place here does not care for the covariance between the covariates that conform $\bm{X}$, as we are only interested on obtaining covariates with similar magnitudes. Also, in practice the standardization is made considering the sample mean and the sample standard deviation of the covariates at hand.
}

\begin{figure}[!ht]
\centering \includegraphics[scale=0.65]{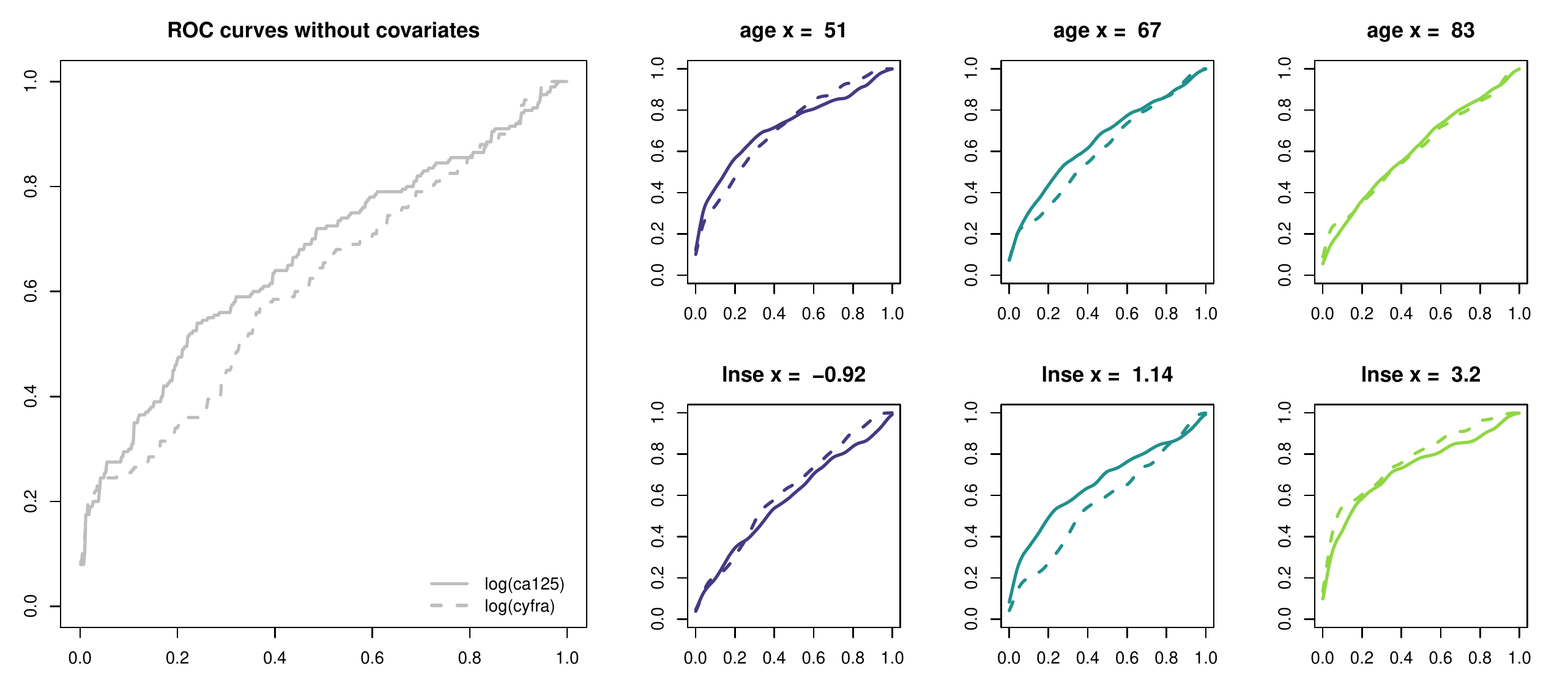}
\caption{ROC curve estimation for both diagnostic variables (\textit{log(ca125)}  and \textit{log(cyfra)}, represented by the solid and the dashed line, respectively) without covariates and conditioned to different values of the covariates \textit{age} and  ${\log(nse)}$.}
\label{fig:graph_ROC_estim}
\end{figure}

We start the analysis of the performance of the two diagnostic markers by comparing their respective ROC curves without taking into account any covariate information. For that matter we use the method proposed by \cite{DeLong1988}. The estimated ROC curves for both markers are depicted in Figure \ref{fig:graph_ROC_estim}. The p-value obtain for that comparison was 0.138. Similar results were obtained when using other ways of comparing ROC without covariates (like \cite{Martinez-Camblor2013a} or \cite{Venkatraman1996}). Thus, we do not find significant differences between the two diagnostic variables in terms of diagnostic accuracy.

Next, we compare the two diagnostic markers taking into account a unidimensional covariate using the test proposed in \cite{Fanjul-Hevia2019} for dependent diagnostic markers. We consider the covariates \textit{age} and ${\log(nse)}$, each one at a time. We test the equality of the ROC curves conditioned to  the values of $\{51,67,83\}$ in the case of \textit{age} and the values of $\{-0.92,1.14,3.27\}$ in the case of ${\log(nse)}$. The corresponding ROC curve for every case is estimated in Figure \ref{fig:graph_ROC_estim}.
For each considered covariate and each value of the covariate we obtain a p-value of the test, summarized in Table~\ref{table:ResCovUni}. The test is made considering two types of statistics, one based on the $L_2$-measure and the other in the Kolmogorov-Smirnov criteria, although both of them yield similar results.
\begin{table}[!t]
\begin{center}
\begin{tabular}{cccc}
\textit{age} &\textbf{51} & \textbf{67} & \textbf{83}  \\ 
\hline
p-values ($L2$) & 0.454 & 0.218 & 0.936
\end{tabular} 
\qquad
\begin{tabular}{cccc}
\textit{ age} &\textbf{51} & \textbf{67} & \textbf{83}  \\ 
\hline
p-values ($KS$) & 0.512 & 0.202 & 0.762
\end{tabular} 

\vskip12pt
\begin{tabular}{cccc}
${\log(nse)}$ &\textbf{-0.92} & \textbf{1.14} & \textbf{3.20}   \\ 
\hline
p-values ($L2$) & 0.844  & {\color{gray}0.012} & 0.470
\end{tabular} 
\qquad 
\begin{tabular}{cccc}
${\log(nse)}$ &\textbf{-0.92} & \textbf{1.14} & \textbf{3.20}  \\ 
\hline
p-values ($KS$) & 0.900 & {\color{gray}0.008} & 0.412
\end{tabular} 
\end{center}
\caption{Results for the comparison of the ROC curves of the diagnostic markers ${\log(ca125)}$ and ${\log(cyfra)}$ when considering a unidimensional covariate, that covariate being the \textit{age} or the ${\log(nse)}$.}
\label{table:ResCovUni}
\end{table}
When comparing the ROC curves conditioned on different values of the \textit{age}, the results are in line with the obtained for the previous case, in which no covariates where taken into account: the equality of the two curves is not rejected. However, when considering the covariate ${\log(nse)}$, we see that for a certain value (1.14) the null hypothesis is rejected (for a significance level of 5$\%$). This matches  the representation of the conditional ROC curves depicted in Figure~\ref{fig:graph_ROC_estim}. 

Finally, we compare the performance of the two diagnostic variables considering the effect of both the \textit{age} and the ${\log(nse)}$ at the same time. This is where we use the methodology proposed in this paper. We test the equality of their respective ROC curves conditioned to nine pairs of values of the two covariates: the ones obtained by making all the possible combinations of $\{51,67,83\}$ and $\{-0.92,1.14,3.27\}$. As before, two different type of statistics were considered: $L_2$ and $KS$ (and once again, the results are similar in both cases). The results obtained are summarized in Table~\ref{table:ResCovMulti}. 
\begin{table}[!t]
\begin{center}
\begin{tabular}{c|ccc}
\backslashbox{\textit{$\log(nse)$}}{\textit{age}} &\textbf{51} & \textbf{67} & \textbf{83} \\ 
\hline 
\textbf{-0.92} & {\color{gray} 0.000}  & {\color{gray} 0.030}  & 0.258  \\ 
\textbf{1.14}  &  0.152 &  0.070  & {\color{gray} 0.004}  \\ 
\textbf{3.20}  & {\color{gray} 0.026}  & 0.056  &  {\color{gray}0.010} \\ 
\end{tabular} 
\qquad 
\begin{tabular}{c|ccc}
\backslashbox{\textit{$\log(nse)$}}{\textit{age}} &\textbf{51} & \textbf{67} & \textbf{83} \\ 
\hline 
\textbf{-0.92} & {\color{gray} 0.004}  & {\color{gray} 0.048}  & 0.424  \\ 
\textbf{1.14}  & 0.212  & 0.050  &  {\color{gray}0.016} \\ 
\textbf{3.20}  & 0.066 & 0.196 &  {\color{gray}0.032} \\ 
\end{tabular} 
\end{center}
\caption{Results for the comparison of the ROC curves of the diagnostic markers \textit{log(ca125)} and \textit{log(cyfra)} when considering  the multidimensional covariate (\textit{age},\textit{$\log(nse)$}).}
\label{table:ResCovMulti}
\end{table}
Note that in this case we did not represent the estimated ROC curves conditioned to the bidimensional covariate $(age, \log(nse))$. This is to stress the fact that, with this methodology, $\widehat{ROC}^{\bm{x}}$ (with $\bm{x}$ bidimensional) does not need to be computed at all.

The obtained p-values show that, depending on the pair of values of the covariate considered, we can find significative differences between the ROC curves of the ${\log(ca125)}$ and the ${\log(cyfra)}$ markers, including pairs of values that when considered separately in the previous test did not rejected the null hypothesis. Likewise, finding differences between the ROC curves conditioned to marginal covariates at certain values does not mean that those differences will be significant when considering the multidimensional covariates (for example, when we conditioned the ROC curves marginally to the value of 1.14 ${\log(nse)}$ we find differences, but when considering both covariates this difference between the ROC curves only remains significant for the \textit{age} of 83).

\section{Discussion}
\label{sec5}

In this work a new non-parametric methodology has been presented for comparing two or more dependent ROC curves conditioned to the value of a continuous multidimensional covariate. This method combines existing techniques for reducing the dimension in goodness-of-fit tests and for estimating and comparing ROC curves conditioned to a one-dimensional covariate.

A simulation study was carried out in order to analyse the practical performance of the test. Two different functions were proposed for the construction of the statistic, the $L_2$ and the $KS$, the second one being a little more conservative. Different correlations between the diagnostic variables and different sample sizes have been considered, including uneven ones without any appreciable effect on the test performance.


Finally, the methodology was illustrated by means of an application to a data set: with this new test it was possible to detect differences on the discriminatory ability of two diagnostic variables conditioned to two different covariates without the need of an estimator of an ROC curve conditioned to a multidimensional covariate. With this application it becomes clear the importance of being able to include the effect of multidimensional covariates to the ROC curves analysis, as different conclusions could be drawn of the comparison of those curves when considering a multidimensional covariate, when considering unidimensional covariates or when excluding the covariates from the study.

\section*{Acknowledgements}

The research of A. Fanjul-Hevia is supported by the Ministerio de Educación, Cultura y Deporte (fellowship FPU14/05316), as well as by the Spanish Ministerio de Educación y Formación Profesional (Mobility Grant EST18/00673).  A. Fanjul-Hevia,  W. González-Manteiga and I. Van Keilegom acknowledge the support from the Spanish Ministerio de Economía, Industria y Competitividad, through grant number and MTM2016-76969-P, which includes support from the European Regional Development Fund (ERDF). 
 J.C. Pardo-Fernández acknowledges financial support from grant MTM2017-89422-P, funded by the Spanish Ministerio de Economía, Industria y Competitividad, the Agencia Estatal de Investigación and the ERDF. I. Van Keilegom is financially supported by the European Research Council (2016-2021, Horizon 2020 / ERC grant agreement No. 694409).
 The Supercomputing Center of Galicia (CESGA) is acknowledged for providing the computational resources that allowed to run most of the simulations. Dr. F. Gude (Unidade de Epidemioloxía Clínica, Hospital Clínico Universitario de Santiago) is thanked for providing the data set analysed in this article.

\section*{Appendix: proofs}


The proofs needed for Lemma~\ref{lemma0} are presented below.

\begin{lemma} \cite{Escanciano2006} or \cite{Cuesta-Albertos2019}:
Given a random variable $Y$ such that $\mathbb{E}|Y|<\infty$,

\begin{eqnarray}
\mathbb{E}[Y|\bm{X}] = 0 \; a.s. \Leftrightarrow \mathbb{E}[Y|\bm{\beta' X}] = 0 \;  a.s. \text{ for any vector $\bm{\beta} \in \mathbb{S}^{d-1}$}.
\label{eq:Esc}
\end{eqnarray}

\end{lemma}

From now on it will be assumed that all projections $\bm{\beta}$ considered satisfy $\bm{\beta} \in \mathbb{S}^{d-1}$.

\begin{lemma}
Let $Y_1, \cdots , Y_K$ be $K$ dependent random variables with cumulative distribution functions $F_1, \ldots, F_K$, respectively, such that $\mathbb{E}|Y_k|<\infty$ for every $k \in\{1,\ldots,K\}$. Let $\bm{X}$ be a multidimensional covariate. Then, given $c_1,\ldots, c_K $,
\begin{eqnarray}
F_1(c_1|\bm{X}) = \cdots = F_K(c_K|\bm{X}) \; a.s.  \Leftrightarrow  F_1(c_1|\bm{\beta'X}) = \cdots = F_K(c_K|\bm{\beta'X}) \; a.s. \; \forall \bm{\beta},
\label{eq:FbetaX}
\end{eqnarray}
with $\bm{\beta} \in \mathbb{S}^{d-1}$.

\begin{proof}
It is proven for $K=2$:
\begin{eqnarray*}
F_1(c_1|\bm{X}) = F_2(c_2|\bm{X}) \; a.s.  &\Leftrightarrow & 
\mathbb{E}[I(Y_1 \leq c_1 )| \bm{X}]  =\mathbb{E}[I(Y_2 \leq c_2) | \bm{X}] \; a.s.  \notag \\
&\stackrel{(*)}{\Leftrightarrow} &  \mathbb{E}[I(Y_1 \leq c_1) - I(Y_2 \leq c_2)| \bm{X}] = 0 \; a.s. \notag \\
& \stackrel{(\ref{eq:Esc})}{\Leftrightarrow}&  \mathbb{E}[I(Y_1 \leq c_1) - I(Y_2 \leq c_2)| \bm{\beta' X}] = 0 \; a.s. \; \forall \bm{\beta} \notag \\
&\Leftrightarrow & 
\mathbb{E}[I(Y_1 \leq c_1 )| \bm{\beta'X}]  =\mathbb{E}[I(Y_2 \leq c_2) | \bm{\beta'X}] \; a.s. \; \forall \bm{\beta}\notag \\
 &\Leftrightarrow & F_1^{\bm{\beta}}(c_1|\bm{\beta'X}) = F_2^{\bm{\beta}}(c_2|\bm{\beta'X}) \; a.s. \; \forall \bm{\beta}, 
\end{eqnarray*}
where $F_i^{\bm{\beta}}(c_i|\bm{\beta'X}) = P(Y_i \leq c_i | \bm{\beta'X} = \bm{\beta'X})$ for $i =1,2$.

Note that in (*) the fact that the random variables are dependent is being used in the sense that they are conditioned to the same covariate $\bm{X}$ (i.e., there is no $X_1$ and $X_2$ as there would be in the independent case).
\end{proof}
\end{lemma}

\begin{mydef}
The \textit{inverted conditional ROC curve (IROC)} is defined as:
\[IROC(p)=1-G(F^{-1} (1-q)), \; q \in (0,1).\]

Related to the previous definition, the \textit{inverted conditional ROC curve ($IROC^x$)}, given the pair $(x^F,x^G)\in R_{X^F}\times R_{X^G}$, can also be defined  as:
\[IROC^{x^G,x^F}(q) = 1-G(F^{-1} (1-q|x^F)|x^G), \; q \in (0,1). \]
\end{mydef}

\begin{lemma} The equality of ROC curves is equivalent to the equality of the inverted ROC curves, i.e.,
\[ROC_1 (p) = \cdots = ROC_K(p) \; \forall p \in (0,1) \; \Leftrightarrow \; IROC_1 (q) = \cdots = IROC_K(q) \; \forall q \in (0,1).\]
Moreover, the same property holds when talking about conditional ROC curves. Given the pair $(x^F,x^G)\in R_{X^F}\times R_{X^G}$ , 
\begin{eqnarray}
ROC_1^{x^F,x^G} (p) = \cdots = ROC_K^{x^F,x^G}(p) \; \forall p \in (0,1) \; \Leftrightarrow \; IROC_1^{x^G,x^F} (q) = \cdots = IROC_K^{x^G,x^F}(q) \; \forall q \in (0,1).
\label{eq:ROCIROC}
\end{eqnarray}
\begin{proof}
It is proven for the unconditional case, and for $K=2$. The conditional case is similar.
\begin{eqnarray*}
ROC_1(p) = ROC_2(p) \; \forall p \in (0,1) &\Leftrightarrow & 
1-F_1(G_1^{-1} (1-p)) = 1-F_2(G_2^{-1} (1-p)) \; \forall p \in (0,1) 
\end{eqnarray*}
Take $q = 1-F_2(G_2^{-1} (1-p))$ (and hence, $q = ROC_2(p) $). $q$ will take all the values in $(0,1)$, and thus, $p = 1-G_2(F_2^{-1}(1-q)) = IROC_2(q)$.

Then,
\begin{eqnarray*}
ROC_1(p) = ROC_2(p) \; \forall p \in (0,1) &\Leftrightarrow & 
1-F_1(G_1^{-1} (1-(1-G_2(F_2^{-1}(1-q)) )) = q \; \forall q \in (0,1) \notag \\
&\Leftrightarrow & 1-G_2(F_2^{-1}(1-q) = 1- G_1(F_1^{-1}(1-q)) \; \forall q \in (0,1)\notag \\
&\Leftrightarrow & IROC_2 (q) = IROC_1(q) \; \forall q \in (0,1).
\end{eqnarray*}
\end{proof}
\end{lemma}

\textbf{Proof of Lemma~\ref{lemma0}}
\begin{proof}
It is proven for $K=2$. For $p \in (0,1)$,
\begin{eqnarray*}
ROC_1^{\bm{x}}(p) &=& ROC_2^{\bm{x}}(p) \; a.s.  \Leftrightarrow \notag \\
&\Leftrightarrow & 
1- F_1(G_1^{-1} (1-p|\bm{x})|\bm{x}) = 1- F_2(G_2^{-1} (1-p|\bm{x})|\bm{x} ) \; a.s. \notag \\
&\Leftrightarrow &
F_1(G_1^{-1} (1-p|\bm{x})|\bm{x}) = F_2(G_2^{-1} (1-p|\bm{x})|\bm{x}) \; a.s. 
\notag \\
&\stackrel{(\ref{eq:FbetaX})}{\Leftrightarrow} & F_1^{\bm{\beta}^F}\left( G_1^{-1} (1-p|\bm{x}) |(\bm{\beta}^F)' \bm{x}\right) =  F_2^{\bm{\beta}^F}\left( G_2^{-1} (1-p|\bm{x} )|(\bm{\beta}^F)'\bm{x}\right) \; a.s. \; \forall \bm{\beta}^F
\notag \\
&\Leftrightarrow & 
ROC_1^{(\bm{\beta}^F)' \bm{x}, \bm{x}}(p) = ROC_2^{(\bm{\beta}^F)' \bm{x} , \bm{x}}(p)  \; a.s. \; \forall \bm{\beta}^F
 \notag \\
& \stackrel{(\ref{eq:ROCIROC})}{\Leftrightarrow}& 
IROC_1^{\bm{x},(\bm{\beta}^F)'\bm{x}}(q) = IROC_2^{\bm{x},(\bm{\beta}^F)' \bm{x}}(q) 
 \; a.s. \; \forall \bm{\beta}^F \;\text{ for } q \in (0,1)
\notag \\
&\Leftrightarrow & 
G_1((F_1^{\bm{\beta}^F})^{-1}(1-q|(\bm{\beta}^F)' \bm{x})| \bm{x}) = G_2((F_2^{\beta^F})^{-1}(1-q|(\bm{\beta}^F)' \bm{x})| \bm{x})
 \; a.s. \; \forall \bm{\beta}^F
\notag \\
&\stackrel{(\ref{eq:FbetaX})}{\Leftrightarrow}&
G_1^{\bm{\beta}^G}((F_1^{\bm{\beta}^F})^{-1}(1-q|(\bm{\beta}^F)' \bm{x})| (\bm{\beta}^G)' \bm{x}) = G_2^{\bm{\beta}^G}((F_2^{\bm{\beta}^F})^{-1}(1-q|(\bm{\beta}^F)' \bm{x})| (\bm{\beta}^G)'\bm{x})
\; a.s. \; \forall \bm{\beta}^F, \bm{\beta}^G
\notag \\
&\Leftrightarrow & 
IROC_1^{(\bm{\beta}^G)' \bm{x},(\bm{\beta}^F)'\bm{x}}(q) = IROC_2^{(\bm{\beta}^G)' \bm{x},(\bm{\beta}^F)'\bm{x}}(q) 
\; a.s. \; \forall \bm{\beta}^F, \bm{\beta}^G
\notag \\
& \stackrel{(\ref{eq:ROCIROC})}{\Leftrightarrow}&  
ROC_1^{(\bm{\beta}^F)'\bm{x},(\bm{\beta}^G)'\bm{x}}(\tilde{p}) = ROC_2^{(\bm{\beta}^F)'\bm{x},(\bm{\beta}^G)'\bm{x}}(\tilde{p}) 
\; a.s. \; \forall \bm{\beta}^F, \bm{\beta}^G  \; \text{for $\tilde{p}\in (0,1)$},
\end{eqnarray*}
where $F_1^{\bm{\beta}^F} \left(c|(\bm{\beta}^F)'\bm{x}\right) = P\left(Y_1^F \leq c |(\bm{\beta}^F)'\bm{X}^F = (\bm{\beta}^F)'\bm{x}\right) $, $F_2^{\bm{\beta}^F} \left(c|(\bm{\beta}^F)'\bm{x}\right) = P\left(Y_2^F \leq c |(\bm{\beta}^F)'\bm{X}^F = (\bm{\beta}^F)'\bm{x}\right) $, $G_1^{\bm{\beta}^G} \left(c|(\bm{\beta}^G)'\bm{x}\right) = P\left(Y_1^G \leq c |(\bm{\beta}^G)'\bm{X}^G = (\bm{\beta}^G)'\bm{x}\right) $, $G_2^{\bm{\beta}^G} \left(c|(\bm{\beta}^G)'\bm{x}\right) = P\left(Y_2^G \leq c |(\bm{\beta}^G)'\bm{X}^G = (\bm{\beta}^G)'\bm{x}\right)$.

\end{proof}


\bibliographystyle{apalike}
\bibliography{biblio} 


\end{document}